\documentclass{article}

\usepackage[usenames,dvipsnames,svgnames,x11names]{xcolor}

\newcommand{\ysnoted}[1]{} 

\usepackage[normalem]{ulem} 

\usepackage{textcomp}

\usepackage[nocompress]{cite}
\usepackage{lipsum}
\let\OLDthebibliography\thebibliography
\renewcommand\thebibliography[1]{
  \OLDthebibliography{#1}
  \setlength{\parskip}{0pt}
  \setlength{\itemsep}{0pt plus 0.3ex}
}

\usepackage{listings}

\lstdefinelanguage{XML}
{
basicstyle=\ttfamily\footnotesize,
  morestring=[b]",
  moredelim=[s][\bfseries\color{Maroon}]{<}{\ },
  moredelim=[s][\bfseries\color{Maroon}]{</}{>},
  moredelim=[l][\bfseries\color{Maroon}]{/>},
  moredelim=[l][\bfseries\color{Maroon}]{>},
  morecomment=[s]{<?}{?>},
  morecomment=[s]{<!--}{-->},
  commentstyle=\color{gray},
  stringstyle=\color{blue},
  identifierstyle=\color{red}
}
\usepackage{moreverb}

\usepackage[nounderscore]{syntax}

\usepackage[pdftex]{graphicx}
\graphicspath{{./figures/}}
\DeclareGraphicsExtensions{.pdf}
\usepackage[cmex10]{amsmath}
\usepackage{amssymb}
\usepackage{mathtools}
\usepackage{amsthm}
\usepackage{amsfonts}
\usepackage[mathscr]{euscript}

\newcommand{\textem}[1]{\text{\emph{#1}}} 

\usepackage{subfig}
\usepackage{algorithmicx}
\usepackage{algpseudocode}
\usepackage[ruled]{algorithm}
\definecolor{light-gray}{gray}{0.75}
\algrenewcommand{\algorithmiccomment}[1]{\hskip3em{{\footnotesize \textcolor{light-gray}{$\blacktriangleright$}}} #1}
\usepackage{multirow} 
\usepackage{rotating} 
\usepackage{booktabs} 
\usepackage{colortbl} 
\usepackage{tablefootnote} 

\usepackage[pdftex,colorlinks=true,urlcolor=blue,citecolor=blue]{hyperref}

\usepackage{xspace}

\usepackage{enumitem}
\newcommand{\para}[1]{\noindent\textbf{#1.~}}

\usepackage{blindtext}

\def\BibTeX{{\rm B\kern-.05em{\sc i\kern-.025em b}\kern-.08em
    T\kern-.1667em\lower.7ex\hbox{E}\kern-.125emX}}
\begin{document}
%
\title{Collaborative Reuse of Streaming Dataflows in IoT Applications}

\author{Shilpa Chaturvedi, Sahil Tyagi and Yogesh Simmhan \\
\normalsize{\emph{Department of Computational and Data Sciences}}\\
\normalsize{\emph{Indian Institute of Science (IISc), Bangalore 560012, India}}\\
\normalsize{\emph{Email: shilpa@grads.cds.iisc.ac.in, simmhan@cds.iisc.ac.in}}}


%


\date{}
\maketitle

\begin{abstract}
Distributed Stream Processing Systems (DSPS) like Apache Storm and Spark Streaming enable composition of continuous dataflows that execute persistently over data streams. They are used by Internet of Things (IoT) applications to analyze sensor data from Smart City cyber-infrastructure, and make active utility management decisions. As the ecosystem of such IoT applications that leverage shared urban sensor streams continue to grow, applications will perform duplicate pre-processing and analytics tasks. This offers the opportunity to collaboratively reuse the outputs of overlapping dataflows, thereby improving the resource efficiency. In this paper, we propose \emph{dataflow reuse algorithms} that given a submitted dataflow, identifies the intersection of reusable tasks and streams from a collection of running dataflows to form a \emph{merged dataflow}. Similar algorithms to unmerge dataflows when they are removed are also proposed. We implement these algorithms for the popular Apache Storm DSPS, and validate their performance and resource savings for 35 synthetic dataflows based on public OPMW workflows with diverse arrival and departure distributions, and on 21 real IoT dataflows from RIoTBench. 
We see that our Reuse algorithms reduce the count of running tasks by $38-46\%$ for the two workloads, and a reduction in cumulative CPU usage of $36-51\%$, that can result in real cost savings on Cloud resources.
\end{abstract}


%

\section{Introduction}
One of the fast growing sources of data is from \emph{Internet of Things (IoT)} deployments, where sensors and actuators collect observational data from and enact control signals on physical and virtual infrastructure~\cite{Perera:2014}. While consumer IoT devices like FitBit and Nest are popular, Smart Cities offer a canonical use of IoT technologies to provide effective citizen services, and improve the efficiency of the utility infrastructure. Examples of these Cyber-Physical Systems (CPS) include \emph{smart power grids} where real-time load measurements from consumers help with demand-response optimization~\cite{aman:tkde:2015}, and \emph{intelligent transportation} where street sensors and camera feeds are used to manage traffic lights, transit frequency, and pricing~\cite{biem2010ibm}.

Smart City deployments make streaming data available from possibly millions of sensors, and need to analyze and process them in near real-time to make decisions or provide services. \emph{Distributed Stream Processing Systems (DSPS)} offer a \emph{fast data} platform to compose \emph{continuous dataflow} applications 
that execute constantly over one or more streams. DSPS like \emph{Apache Storm, Flink} and \emph{Spark Streaming}~\cite{storm,flink, sparkstreaming} are designed to scale-out across commodity clusters and Cloud Virtual Machines (VMs), and operate on $1000$'s of $messages/sec$. They are commonly used to compose IoT and Smart City applications hosted on the Cloud, and access sensor streams pulled from the edge into the data-center~\cite{giang2015developing,simmhan:cise}. 

\para{Motivation} 
As Smart City installations expand, 
thousands of public observation streams on traffic, pollution, weather, etc. 
from diverse domains will be available 
for integration and analysis. 
One can expect an explosion of innovative services and ``apps'' that perform online analytics over these 
streams, 
even personalize it for individuals. 
E.g., an app may correlate weather observation streams (\emph{turning cloudy}) with power grid generation streams (\emph{solar output drop}) to predict when surge-pricing might be triggered by the utility to offset demand. This can help users (or their digital agents) schedule, say, a recharge of their electric vehicle or their smart washing machine. 

Cloud-hosted DSPS will form the scalable analytics engine for composing and executing these continuous dataflows, collocated with the data streams. At the same time, there will be duplication of tasks by the numerous dataflows that operate on these shared streams, which may each perform similar data pre-processing (\emph{parsing, reformat, unit conversion}), quality checks (\emph{cleaning, outlier detection, interpolation}), and even analytics (\emph{ARIMA time-series predictions, moving window averages})~\cite{riotbench,nastic2013patricia}. This offers the opportunity to reuse parts of the logic among different dataflows to avoid recomputation, thereby reducing the costs for using Cloud resources for app developers and end-users, and the time to deployment as well.

%
\para{Gaps} Such scenarios are common in eScience applications where datasets and workflows are reused. Scientific projects often make Level 1/2/3 datasets, which have been pre-processed to different degrees using standard routines, available to their user groups. Similarly, repositories like \emph{myExperiment} and \emph{OPMW} allow the definition and reuse of scientific workflows by the broader community~\cite{myexperiment,opmw}. Provenance collected from workflow runs have also been leveraged for data and workflow reuse~\cite{davidson2008provenance,simmhan:sigmodrec:2005}. Even \emph{Apache Spark} uses lineage to avoid recomputing RDDs~\cite{Zaharia:2012:RDD:2228298.2228301}. Others have examined stream reuse in wide area networks~\cite{4620109}.

While related, the problem we address differs from these prior works. Reuse of workflows and their outputs happens after their execution. We instead focus on streaming applications from diverse users that are actively running and generating transient data streams. This requires a greater awareness of the platform runtime, and is performance sensitive. 
The IoT community is nascent, spanning startups, citizen scientists, and utilities. 
While it is premature for formal dataflow standards to be adopted, fast data platforms like Apache Storm and Flink, evolving IoT libraries, and public Clouds offer the lowest common denominator~\cite{riotbench}. We leverage these.


\para{Contributions} In this paper, we make the following specific contributions:
\begin{enumerate}[leftmargin=0.4cm,itemindent=0cm,labelwidth=0.4cm,labelsep=0cm,align=left]
\item We motivate (\S~\ref{sec:problem}) and formally define (\S~\ref{sec:definition}) the problem of streaming dataflow reuse, including the \emph{equivalence} between tasks present in dataflows.
\item We propose algorithms for \emph{merging} a submitted streaming dataflow with deployed dataflows at specific points of equivalence, and similarly, \emph{unmerging} a merged dataflow when it is removed, while guaranteeing their output stream consistency, in \S~\ref{sec:solution}.
\item We implement our reuse algorithms in \emph{Apache Storm}, and \emph{validate} it for real and synthetic Smart Utility applications and public OPMW workflows~(\S~\ref{sec:results}).
\end{enumerate}

We also review related literature in \S~\ref{sec:related}, and present our conclusions and future work in \S~\ref{sec:conclusion}.

\section{Problem Description}
\label{sec:problem}


Continuous or \emph{streaming dataflows} are composed as a \emph{Directed Acyclic Graph (DAG)}, where vertices are user logic or \emph{tasks} and directed edges are \emph{streams} that transfer opaque \emph{events} between the output of a task to the input of another downstream task. Tasks execute once per input event to generate zero or more output events, with the ability to aggregate local state and operate on multiple input events. Such dataflows, once \emph{deployed} onto a DSPS, are execute continuously on their input stream(s) till \emph{undeployed}.

Such streaming dataflows have been used to compose IoT applications which execute on the Cloud and operate over input streams from \emph{physical sensors} (sometimes online feeds) that are available publicly~\cite{biem2010ibm,riotbench}. The dataflows themselves may publish output streams, or have a sink task that persists the output events to storage. The output streams from each task in the dataflow can also be considered as an intermediate stream that has been partially processed through the preceding dataflow tasks. We refer to these output and intermediate streams as \emph{derived streams} that have been processed, in contrast to the \emph{raw streams} from sensors. 

DSPS like Apache Storm and Flink can run multiple concurrently dataflows on a commodity cluster or Cloud VMs. Dataflows submitted to a Storm installation execute independently on a common set of hosts configured for the Storm cluster. Tasks from multiple dataflows can be collocated on the same machine
, but there is no implicit sharing of events or tasks between different dataflows. 

IoT dataflows that use the same raw stream(s) as input(s) are likely to have similar pre-processing or even analytics tasks. This is particularly so when Smart Cities make many observation streams public for startups and citizen scientists to design novel applications for the residents, running on public Clouds, or private city-hosted Clouds. As a result, it is likely that dataflows with significant overlaps between them will run independently, thus duplicating their efforts. 

\begin{figure}[t!]
  \centering
	\includegraphics[width=0.99\columnwidth]{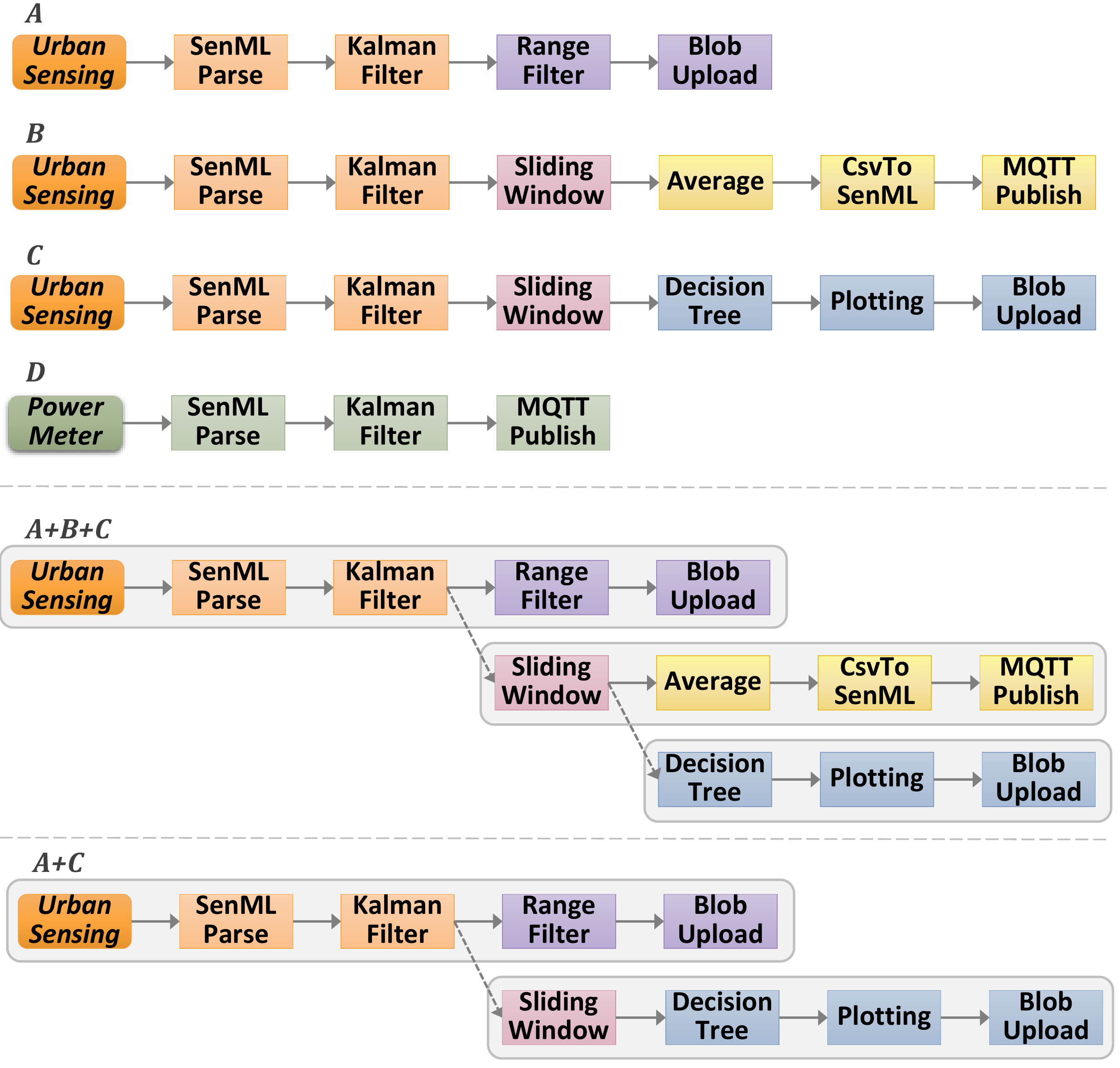}
	\caption{Illustration of dataflows being merged for reuse on submission, and unmerged on removal.}
    \label{fig:sample}
\vspace{-0.1in}
\end{figure}
Fig.~\ref{fig:sample} illustrates such a scenario where dataflows $A, B, C$ and $D$ are performing \emph{Extract-Transform-Load (ETL)} and \emph{Statistical Summarization (STATS)} on two streams, from urban sensing and smart power meters~\cite{riotbench}. The dataflows differ in overall structure but share similar prefix tasks. E.g., Dataflow $A, B$ and $C$ share the raw stream source and the next two tasks, while $B$ and $C$ share an additional third task. As a result, these three dataflows can be ``merged'' into one dataflow, $A+B+C$, where $B$ reuses a copy of the derived stream from $A$'s \emph{Kalman Filter} output, and likewise $C$ reuses a copy of the derived stream from $B$'s \emph{Sliding Window} output. This achieves the same result as running the three independently, but avoids duplicate execution of the prefix tasks. We see that dataflow $D$ has an overlap with $A$ but the source stream is different, and hence they cannot be merged. Similarly, when dataflow $B$ is undeployed, then an ``unmerge'' should bring it to $A+C$. 

Each of these dataflows may be owned by a different user who is part of the IoT community, and they collaboratively wish to reuse the dataflows to reduce their costs due to redundant computation. While these examples show simple sequential dataflows being merged and unmerged, there can be more complex scenarios where DAGs have forks and joins, dataflows are added and removed in arbitrary order, and tasks may have additional configuration parameters. While dataflows make the composition simple, manually identifying the overlaps with existing dataflows for cost-efficient execution infeasible 
in an active Smart City ecosystem with hundreds of users and their applications.

In this paper, we explore algorithms to transparently reuse derived streams in submitted dataflows to reduce resource utilization while guaranteeing that the outputs of the dataflows are identical to the original ones, even when the reused dataflows are removed. There are specific challenges on \emph{correctness and efficiency} that our solution must address.
\begin{itemize}[leftmargin=0.5cm,itemindent=0cm,labelwidth=0.5cm,labelsep=0cm,align=left]
\item  We need to \emph{automatically identify} the derived streams in existing dataflows that offer the logical equivalent of a stream in the submitted dataflow. This requires checking that the \emph{ancestors} (causal chain, provenance) of the derived stream matches the one in the new dataflow. The raw input stream(s), the task types and their configurations must identical. 
\item We also need to ensure that this \emph{reuse is maximal}, and as far downstream as possible, to take best advantage of the deployed dataflows. 
\item We should support the reuse of \emph{multiple derived streams from different dataflows} by the same incoming dataflow. 
\item When a dataflow is removed, the \emph{unmerging} should retain the correctness of the remaining (merged) dataflows while also minimizing the disruption to existing applications. Dependencies should be accurately resolved. 
\ysnoted{\item Book-keeping should track the degree of reuse between dataflows to allow \emph{accurate billing} of Cloud resources for owners of the dataflow.}
\end{itemize}

Next, we formalize this problem and propose dataflow merge and unmerge algorithms to meet these requirements.




\section{Problem Formulation}
\label{sec:definition}

\subsection{Tasks, Streams and Dataflows}
An \textbf{event} is a discrete unit of data that is uniquely identified by an \emph{event id}, and has a \emph{payload} whose contents are opaque to the platform. An \textbf{abstract task} $\tau=\langle\textem{type},\textem{config} \rangle$ is a user-defined logic, as determined by its \emph{type}, which consumes and operates on one event at a time, and may generate zero or more events for each event consumed. The behavior of the user logic is controlled by parameters specified in a \emph{config} property for the task, such as the window size for an aggregation task or the NoSQL URL for an event storage task. A \textbf{stream} is a logical channel to transfer events generated from a task to a destination task for consumption.

Let $\mathcal{T} = \{ \tau \}$ be the universal set of all distinct abstract tasks. Two abstract tasks are identical if their type and their config are the same, \[ \tau_i = \tau_j \implies \tau_i.type = \tau_j.type \land \tau_i.config =  \tau_j.config \] 
%
%
%
%
\textbf{Source tasks} and \textbf{sink tasks} are special abstract tasks that serve solely as generators and consumers of event in streams, respectively. A source task does not consume an input stream, but produces (raw) events on its output stream based on its internal logic (e.g., read from a physical sensor), while a sink task consumes an input stream but does not produce an output stream (e.g., persist to a NoSQL database).   
Their \emph{type} uniquely identifies the logical name of the source or sink while their \emph{config} has a constant value of \textsc{`source'} or \textsc{`sink'}. The sets $\mathcal{R} \subset \mathcal{T}$ and $\mathcal{N} \subset \mathcal{T}$ are the universal set of source and sink tasks, with $\mathcal{R} \cap \mathcal{N} = \varnothing$.



Users compose streaming applications as a \textbf{dataflow} defined as a Directed Acyclic Graph (DAG), $D = \langle \mathbb{T}, \mathbb{S} \rangle$, where $\mathbb{T}=\{t_1,...,t_n\}$ is the set of $n$ \textbf{concrete tasks} (or just ``tasks'') that form the vertices of the DAG, and $\mathbb{S}=\{s_1,...,s_m\}$ is the set of $m$ \emph{streams} that are the edges of the DAG. 
Each concrete task $t_i \in \mathbb{T}$ has an \emph{id} that is globally unique, in addition to matching an abstract task's \emph{type} and \emph{config}, \[ t_i=\langle \textem{id},\textem{type}_p,\textem{config}_q\rangle ~\mid~ \exists \tau=\langle \textem{type}_p,\textem{config}_q\rangle  \in \mathcal{T} \]
The same abstract task can appear multiple times as different concrete tasks in the DAG with different \emph{id}'s. A stream $s_k \in \mathbb{S}$ that transfers output events from an upstream task $t_i$ to the input of a downstream task $t_j$ is defined as, \[ s_k=\langle t_i.id,t_j.id \rangle ~\mid~ t_i,t_j \in \mathbb{T}\]We follow \emph{interleave semantics} when multiple streams are incident on the same task, i.e., the task is executed once for each event that arrives on any of its input streams, and \emph{duplicate semantics} when multiple streams leave the same task, i.e., a copy of each event generated by the task is placed on each of its output streams. Hence, task and stream reuse are interchangeable. Two convenience functions return the upstream and downstream tasks an edge is incident on,
\[up(s) = \{t_i \mid s = \langle t_i.id, t_j.id \rangle \in \mathbb{S},~ t_i, t_j \in \mathbb{T} \} \]
\[down(s) = \{t_j \mid s = \langle t_i.id, t_j.id \rangle \in \mathbb{S},~ t_i, t_j \in \mathbb{T} \} \]

A dataflow has a set of \textbf{input} and \textbf{output tasks} which are the start and the end boundaries of the DAG, and should be part of the universal set of source and sink tasks. $\mathbb{I} = \mathbb{T} \cap \mathcal{R}$ is the set of input tasks that pass the input event stream(s) for processing by the DAG, while $\mathbb{O} = \mathbb{T} \cap \mathcal{N}$ is the set of output tasks that manage the output stream(s) of the DAG.

A DSPS engine continuously executes tasks of the dataflow on distributed resources and orchestrates the event transfer. While our definition makes no assumptions on the runtime characteristics or scheduling, our techniques are well-suited for dataflows executed in a single Cloud data center.

\subsection{Equivalence}

\para{Similarity between Tasks}
Say we have two concrete tasks $t_i$ and $t_j$. They are said to be \emph{type-similar} if $t_i.type = t_j.type$, and denoted as $t_i \overset{\text{T}}{\approx} t_j$. They are said to be \emph{config-similar} if they are type similar, and also $t_i.config = t_j.config$, denoted as $t_i \overset{\text{C}}{\approx} t_j$. The tasks are said to be \emph{identical} if $t_i.id = t_j.id$, and denoted as $t_i = t_j$. Being identical implies that these are the same tasks, and so require that they also be config similar. 
Task similarity is orthogonal to its runtime performance.


\para{Parent of a Task}
For a dataflow $D\langle \mathbb{T},\mathbb{S}\rangle$, we define a \emph{parent function}, $\pi_D: \mathbb{T} \to \mathscr{P}(\mathbb{T})$, that takes a task in the dataflow as input and returns its \emph{parent set}, which is the set of tasks that are the immediate upstream predecessors of the given task in the DAG. The function's range falls in the power set $\mathscr{P}$ of all tasks. There are no parents for the input task(s) to the dataflow. For $t \in \mathbb{T}$, we have: 
\begin{equation*}
\pi_D(t)=\begin{cases}
\big\{ p \mid \exists s \in \mathbb{S}, p \in \mathbb{T},\\
\qquad p = up(s), t = down(s) \big\} & \text{if } t \in \mathbb{T} \setminus \mathbb{I}\\
\varnothing & \text{if } t \in \mathbb{I}  \\
\end{cases}
\end{equation*}

\para{Ancestor Graph}
An \emph{Ancestor Graph} for a task in a dataflow is a DAG formed from the task and all its ancestors, along with the streams that connect these tasks within the original dataflow. 
%
For a task $t \in \mathbb{T}$ for a dataflow $D\langle \mathbb{T},\mathbb{S} \rangle$ we have the \emph{ancestor graph recurrence function}, $\alpha_D(t) \to A\langle \overleftarrow{\mathbb{T}},\overleftarrow{\mathbb{S}} \rangle \mid \overleftarrow{\mathbb{T}} \subset \mathbb{T},\overleftarrow{\mathbb{S}}\subset\mathbb{S}$, defined as: 
\[ \alpha_D(t) = A \langle \{ t \}, \{ s \mid down(s) = t, s \in \mathbb{S} \} \rangle ~\cup~
\bigcup\limits_{p \in \pi_D(t)}\alpha_D(p) \]
Here, we include the current task and its incoming streams in the ancestor graph, recursively apply the ancestor function on the parent set of the task and take the union of the parent's ancestor graph. This will recur till we reach the DAG's input tasks, which do not have parents. The union of two ancestor graphs $A_i\langle \mathbb{T}_i,\mathbb{S}_i \rangle$ and $A_j\langle \mathbb{T}_j,\mathbb{S}_j \rangle$ is, 
\[A_k\langle \mathbb{T}_k,\mathbb{S}_k \rangle = A_i \cup A_j = \langle \mathbb{T}_i \cup \mathbb{T}_j, \mathbb{S}_i \cup \mathbb{S}_j \rangle\]

The ancestor graph for a task indicates the set of operations that were performed on one or more input tasks to the DAG in order to derive the input stream to the task. It is similar to the prospective provenance of the events generated from that task~\cite{davidson2008provenance}. Each ancestor graph is connected and forms a DAG. Every task in the dataflow has a unique ancestor graph, and it contains at least one of the input tasks to that dataflow. In a dataflow with a single sink task, the ancestor graph of the sink task is the entire dataflow. 

\para{Maximal Ancestor Graph Set}
The \emph{ancestor graph set}, $\mathbb{A}$, for a dataflow $D\langle \mathbb{T}, \mathbb{S}\rangle$ is given by, \[ \mathbb{A} = \{ \alpha_D(t) \mid t \in \mathbb{T} \} \]
An ancestor graph $A_j\langle \mathbb{T}_j,\mathbb{S}_j \rangle$ is said to be a \emph{sub-ancestor} of another ancestor graph $A_i\langle \mathbb{T}_i,\mathbb{S}_i \rangle$  if $\mathbb{T}_j \subset \mathbb{T}_i$ and $\mathbb{S}_j  \subset \mathbb{T}_i$, and we say that $A_j \subset A_i$.
A \emph{maximal ancestor graph set}, $\widehat{\mathbb{A}}$, for a given dataflow is the ancestor graph set that only contains ancestor graphs that are not sub-ancestors of any other ancestor graph in that dataflow, 
\[ \widehat{\mathbb{A}} = \Omega(\mathbb{A}) = \{ A \mid A \not\subset A',~~A,A' \in \mathbb{A} \} \]
with the function $\Omega$ returning the maximal ancestor graph for any given set of ancestor graphs. 

Intuitively, the number of ancestor graphs in the maximal ancestor graph set in a given dataflow will match the number of sink tasks for that dataflow. This is because the sink being the most downstream of the tasks in the DAG will not appear in any other ancestor graph besides its own. It will also have the most number of tasks in its ancestor graph.

\para{Task and Ancestor Graph Equivalence} 
If we have $A_i\langle \mathbb{T}_i,\mathbb{S}_i \rangle$ and $A_j\langle \mathbb{T}_j,\mathbb{S}_j \rangle$ as the ancestor graphs for tasks $t_i,t_j \in \mathbb{T}$ in a dataflow $D\langle \mathbb{T}, \mathbb{S}\rangle$, we say that the \emph{ancestors graphs are equivalent}, denoted as $A_1 \leftrightarrow A_2$, if there exists a \emph{bijective} function $\epsilon:\mathbb{T}_i \rightarrow \mathbb{T}_j$, \[\epsilon(t_i') = t_j' \implies t_i' \overset{\text{C}}{\approx} t_j',~t_i' \in \mathbb{T}_i, t_j' \in \mathbb{T}_j\] In other words, for each task in the ancestor graph of $t_i$, there should be a distinct task in the ancestor graph of $t_j$ that is config-similar, and vice versa.


Two tasks $t_i$ and $t_j$ are equivalent, denoted as $t_i \leftrightarrow t_j$ if they are config-similar and their ancestor graphs are equivalent. If two tasks are equivalent, then both their output streams are identical, and one can replace the other.





\para{De-Duplicated DAG (De-dup DAG)}
A \emph{De-Duplicated DAG} $D\langle \mathbb{T}, \mathbb{S} \rangle$ 
is one in which there exists no two task $t_i,t_j \in \mathbb{T}$ that are equivalent. Each dataflow submitted by the user for execution should be a de-dup DAG.

\para{Disjoint and Overlapping DAGs}
Two dataflows $D_i\langle \mathbb{T}_i,\mathbb{S}_i \rangle$ and $D_j \langle \mathbb{T}_j,\mathbb{S}_j \rangle$ are said to be \emph{disjoint}, denoted as $D_i \nleftrightarrow D_j$, if they do not have any tasks between them that are equivalent, \[ D_i \nleftrightarrow D_j \implies \not\exists t_i \in \mathbb{T}_i, t_j \in \mathbb{T}_j \mid t_i \leftrightarrow t_j \]
Disjoint dataflows have no tasks that are mutually reusable. Dataflows that are not disjoint are called \emph{overlapping}.

\para{Ancestor Intersection of DAGs}
We define the \emph{ancestor intersection} of two DAGs, given as a function $\Lambda(D_i,D_j)$, as the set of ancestor graphs for tasks in each of the DAGs that are ancestor equivalent. WLOG, we choose the ancestor graph from the task in the first DAG for inclusion in the intersection set. Given $D_i\langle \mathbb{T}_i,\mathbb{S}_i \rangle$ and $D_j \langle \mathbb{T}_j,\mathbb{S}_j \rangle$, we have their ancestor intersection as:
\[\Lambda(D_i,D_j) = \{ \alpha_{D_i}(t_i) \mid t_i \leftrightarrow t_j ~\forall~ t_i \in \mathbb{T}_i, t_j \in \mathbb{T}_j \} \]
The ancestor intersection of disjoint DAGs is an empty set.

The \emph{maximal ancestor intersection} finds the maximal set from the returned set of intersecting ancestor graphs, \[ \widehat{\Lambda}(D_i,D_j) = \Omega(\Lambda(D_i,D_j)) \]
The maximal ancestor intersection indicates the largest set of equivalent tasks in the two dataflows, and offers an upper bound on the tasks that can be reused.

\subsection{Problem Definition}
We distinguish between \emph{submitted DAGs}, that are provided by users for deployment and execution, and \emph{running DAGs} 
that are actually deployed and executing in the DSPS. Our goal is to optimize the set of running DAGs for the given set of submitted DAGs while ensuring the outputs provided to the users by the running DAGs are indistinguishable from those of the submitted DAGs. We also wish to do this dynamically, as the set of submitted DAGs changes, i.e., DAGs are submitted and removed by users over time.

Say, we have a set of $m$ disjoint and de-dup DAGs, $\overline{\mathbb{D}}=\{ \overline{D}_i\langle \overline{\mathbb{T}}_i, \overline{\mathbb{S}}_i \rangle \}$ that are currently \emph{running}, and together represent a collection of $n \ge m$ de-dup DAGs, $\mathbb{D}=\{ D_j\langle \mathbb{T}_j,\mathbb{S}_j \rangle \}$, that were \emph{submitted} by users for deployment. The following two \textbf{constraints} hold for the system: 
\begin{enumerate}[leftmargin=0.5cm,itemindent=0cm,labelwidth=0.5cm,labelsep=0cm,align=left]
\item \para{Sink Task Coverage} For each sink task $t_p$ present in the dataflows $\mathbb{D}$ that were submitted by the users, there exists some task $t_q$ in the running dataflows $\overline{\mathbb{D}}$ that has the same ancestor equivalence,
\begin{equation}\label{cons:cov}
 \forall t_p \in \mathcal{N} \cap \mathbb{T}_j,~~\exists t_q \in \overline{\mathbb{T}}_i \mid t_p \leftrightarrow t_q
\end{equation}
\item \para{Task and Stream Minimization} The running dataflows $\overline{\mathbb{D}}$ must be disjoint and de-dup DAGs. Each task $t_q$ and stream $s_r$ in the running dataflows must appear in the ancestor graph for some sink task $t_p$ in the submitted dataflows $\mathbb{D}$,
\begin{align}\label{cons:min}
\forall t_q \in \overline{\mathbb{T}}_i, s_r \in  \overline{\mathbb{S}}_i~~&\exists t_p \in \mathcal{N} \cap \mathbb{T}_j \mid t_q \in \overleftarrow{\mathbb{T}_p} \land s_r \in \overleftarrow{\mathbb{S}_p} \nonumber\\
& \text{ where } A_p\langle \overleftarrow{\mathbb{T}_p},\overleftarrow{\mathbb{S}_p} \rangle  = \alpha_{D_j}(t_p)
\end{align}
\end{enumerate}

Here, the first constraint guarantees that there is an equivalent task in the running dataflows for every output task in each dataflow submitted by the user. This means that the running dataflows can produce the identical output streams as the submitted dataflows. The second constraint ensures that there are no more running tasks, and streams connecting them, than what is absolutely required to satisfy the equivalence with output tasks in the submitted dataflows. This, coupled with the running dataflows being disjoint, ensures that we execute the least number of tasks required while maximizing reuse. Given this, our \textbf{problems} are,
\begin{enumerate}[leftmargin=0.5cm,itemindent=0cm,labelwidth=0.5cm,labelsep=0cm,align=left]
\item \para{Merging DAGs} When a new de-dup DAG $D_n$ is submitted by a user, update the set of running DAGs $\overline{\mathbb{D}}$ such that Constraints~\ref{cons:cov} and \ref{cons:min} hold for the new set of submitted DAGs $\mathbb{D} \cup D_n$, while ensuring 
  that tasks equivalent to the output tasks of $D_n$ are present in $\overline{\mathbb{D}}$.
%
\item \para{Unmerging DAGs} When a dataflow $D_r \in \mathbb{D}$ that was earlier submitted is now requested to be removed, update the set of running DAGs $\overline{\mathbb{D}}$ such that Constraints~\ref{cons:cov} and \ref{cons:min} hold for the new set of submitted DAGs $\mathbb{D} \setminus D_r$.
\end{enumerate}

\ysnoted{Later, give weightage to the task intensity than just the two constraints}

\section{Merging and Unmerging Dataflows}
\label{sec:solution}

 
\subsection{Merging Algorithm}
When a dataflow is submitted by the user, we need to check if the dataflow is overlapping with any of the running dataflows. If not, then there is no possibility of reuse and the submitted dataflow has to be run independently. If there are overlaps with one or more running dataflows, then we need to identify the overlapping tasks and streams that will be reused. We should also locate the non-overlapping parts of the submitted dataflow that will have to be run afresh, but connected to the upstream tasks being reused. 

Each running dataflow is disjoint with the other running dataflows. This means that they do not share any source tasks between them. Multiple running dataflows can be reused by the same submitted dataflow if it has multiple source tasks (and optionally their successors) that are present in different running dataflows. In this case, these running dataflows will be connected and merged with the new (non-overlapping) tasks and streams that are instantiated for the submitted dataflow. Hence, multiple running DAGs will merge into a single running DAG, along with the newly created tasks and streams. 

We also need to identify the tasks in the running dataflow that correspond to the sink tasks in the submitted dataflow so that the user knows where the output of their dataflow is incident. This should also be maintained for dataflows submitted earlier, when a merge of running dataflows happens. We next discuss specifics of these various operations required for merging a submitted DAG with running ones.

\para{Identifying Overlaps} Say $D_n\langle \mathbb{T}_n, \mathbb{S}_n \rangle $ is the \emph{newly submitted DAG} and $\overline{\mathbb{D}}$ is the \emph{set of currently running DAGs}. We need to identify $\overline{D_i}\langle \overline{\mathbb{T}_i}, \overline{\mathbb{S}_i} \rangle \in \overline{\mathbb{D}} $ that are not disjoint with dataflow $D_n$. While one can test the ancestor equivalence for every pair of tasks in the submitted DAG and the running ones, this will be costly. Instead, we prune this search-space based on the intuition that running DAGs that share a source task with the submitted DAG have at least a minimal overlap, and hence will overlap and be merged together. In contrast, running DAGs that do not have a source task overlap with the submitted one will be disjoint with the submitted DAG. The set of overlapping DAGs is thus,
%
\[ \mathbb{Y} = \{\overline{D_i} : \overline{\mathbb{T}_i} \cap \mathbb{T}_n \cap \mathcal{R} \ne \varnothing   \} \]

\para{Merging and Reusing Overlaps}
 If $|\mathbb{Y}| \ge 1$, we construct a new \emph{merged DAG} $\overline{D_m}\langle \overline{\mathbb{T}_m} ,\overline{\mathbb{S}_m} \rangle$ from these overlapping DAGs by first performing a union of tasks and streams,
\[ \overline{\mathbb{T}_m} = \bigcup\limits_{\overline{D_i}\langle \overline{\mathbb{T}_i}, \overline{\mathbb{S}_i} \rangle \in \mathbb{Y}}\overline{\mathbb{T}_i} \qquad\qquad \overline{\mathbb{S}_m} = \bigcup\limits_{\overline{D_i}\langle \overline{\mathbb{T}_i}, \overline{\mathbb{S}_i} \rangle \in \mathbb{Y}}\overline{\mathbb{S}_i} \]

Now, we identify the parts of the submitted DAG $D_n$ that overlap and are present in this partially merged dataflow $\overline{D_m}$ by examining their maximal ancestor graph set,
\[ \widehat{\mathbb{A}} = \widehat{\Lambda}(\overline{D_m}, D_n) = \Omega(\Lambda(\overline{D_m}, D_n)) \] 
This set of disjoint and maximal ancestor graphs 
contain the set of tasks $\mathbb{T}_o$ and edges $\mathbb{S}_o$ of the new DAG that are already present in the running DAGs and can be reused, 
\[ \mathbb{T}_o =  \bigcup\limits_{A_k\langle \overleftarrow{\mathbb{T}}_k,\overleftarrow{\mathbb{S}}_k  \rangle \in \widehat{\mathbb{A}}} \overleftarrow{\mathbb{T}}_{k} \qquad \qquad \mathbb{S}_o = \bigcup\limits_{A_k\langle \overleftarrow{\mathbb{T}}_k,\overleftarrow{\mathbb{S}}_k  \rangle \in \widehat{\mathbb{A}}} \overleftarrow{\mathbb{S}}_{k} \]

\para{Including Non-overlapping Tasks}
Then we find the parts of the new dataflow that cannot be reused from the running DAGs, and have to be newly instantiated. Say, $\mathbb{T}_x = \mathbb{T}_n \setminus \mathbb{T}_o$ are the set of new non-overlapping tasks to be created. The new streams $\mathbb{S}_x = \mathbb{S}_{x}^{*} \cup \mathbb{S}_{x}^{+}$ will include those streams that connect tasks fully within $\mathbb{T}_x$, and the boundary streams that link the reused tasks with non-overlapping ones that will be down-stream. These are given by.   
\begin{align*}
\mathbb{S}_{x}^{*} =& \{s_n \mid up(s_n), down(s_n) \not\in \mathbb{T}_o\}& \\ 
\mathbb{S}_{x}^{+} =&  \{  s_n \mid up(s_n) \in \mathbb{T}_o, down(s_n) \not\in \mathbb{T}_o  \}, & \forall s_n \in \mathbb{S}_n
\end{align*}
\ysnoted{We need to fix the latter part in above equation}
$\mathbb{T}_x$ and $\mathbb{S}_x$ are the new tasks and streams that have to started and connected. We then merge these new entities with the merged DAG $D_m$ that we are constructing to get, $\overline{\mathbb{T}_m} = \overline{\mathbb{T}_m} \cup \mathbb{T}_x$ and $\overline{\mathbb{S}_m} = \overline{\mathbb{S}_m} \cup \mathbb{S}_x$. We can then replace the overlapping DAGs in $\mathbb{Y}$ with this newly merged DAG to get the updated set of running DAGs, $\overline{\mathbb{D}} = \overline{\mathbb{D}} \setminus \mathbb{Y} \cup \{ \overline{D_m} \}$. We also add the user's DAG to the set of submitted DAGs for book-keeping, $\mathbb{D} = \mathbb{D} \cup D_n$.

\ysnoted{\para{Identifying Output Tasks}}

\subsection{Unmerging Algorithm}
Users can request an earlier submitted dataflow for removal from the system, and these request can come in any arbitrary order, irrespective of the order of submitting the dataflows. When a dataflow is requested for removal, we need to first identify the running (merged) DAG that contains this dataflow -- the dataflow being removed will be present in only one running DAG due to the merge operations. We then determine the tasks and streams in this running DAG that can be removed such that the correctness of other submitted dataflows that continue to remain in the system is not affected. As part of this operation, a single running DAG may be unmerged into multiple DAGs as the components may get disconnected. These will have to be identified. The upper bound on the number of DAGs that will be unmerged is the number of source tasks that are present in the dataflow being removed. 

Let $\Delta : \overline{\mathbb{D}} \to \mathscr{P}(\mathbb{D})$ be a \emph{decomposition function} that maps from a running (merged) DAG to a set of submitted DAGs it supports, and similarly $\Phi : \mathbb{D} \to \overline{\mathbb{D}}$ be an \emph{inverse mapping function} that given a submitted DAG, return the running (merged) DAG that it is contained in. These can be maintained as the merge algorithm is being performed. 

When a DAG, $D_r\langle \mathbb{T}_r, \mathbb{S}_r \rangle \in \mathbb{D}$ is being removed, the running DAG that contains this dataflow and will be affected is, $\overline{D_i} \langle \overline{\mathbb{T}_i}, \overline{\mathbb{S}_i} \rangle = \Phi(D_r)$. Let the set of dataflows that will be continue to be supported by tasks and streams in this running DAG be $\mathbb{D}_s = \Delta(\overline{D_i}) \setminus D_r$. We need to identify the tasks and streams in the running DAG that must be terminated. These tasks will not appear in the ancestor graph of any of the remaining dataflows $\mathbb{D}_s$ that are supported by the merged DAG.
In particular, let the ancestor graphs for the output tasks in the DAGs that remaining be,
\begin{align*}
\mathbb{A} = \{ A_s \mid & A_s = \alpha_{D_k}(t_p), 
\forall t_p \in \mathbb{T}_k \cap \mathcal{N}, ~D_k \langle \mathbb{T}_k, \mathbb{S}_k \rangle \in \mathbb{D}_s \}
\end{align*}
So tasks in this running DAG that can be terminated are those that do not appear in this ancestor graph set,
\[ \mathbb{T}_t = \{ t_q \mid t_q \in \overline{\mathbb{T}_i},~ t_q \not\in \overleftarrow{\mathbb{T}_p}, A_p\langle \overleftarrow{\mathbb{T}_p},\overleftarrow{\mathbb{S}_p} \rangle \in \mathbb{A} \} \]
and the streams that can be disconnected will overlap with these tasks being terminated,
\[ \mathbb{S}_t = \{ s \mid s \in \overline{\mathbb{S}_i},~up(s) = t \lor down(s) = t,  t \in \mathbb{T}_t \} \]

The running DAG $\overline{\mathbb{D}_i}$ will now reduce to $\overline{\mathbb{D}_j}\langle \overline{\mathbb{T}_j}, \overline{\mathbb{S}_j} \rangle$, where $\overline{\mathbb{T}_j} = \overline{\mathbb{T}_i} \setminus \mathbb{T}_t$ and $\overline{\mathbb{S}_j} = \overline{\mathbb{S}_i} \setminus \mathbb{S}_t$.
Next, we need to identify the distinct DAGs that may be formed from unmerging of $\overline{\mathbb{D}_j}$ if the DAG separates into multiple connected components due to stream edges being removed. These can be found by performing an incremental forward traversal from each of the input tasks that are retained in the DAG, $\overline{\mathbb{T}_j} \cap \mathcal{R}$, and forming a new DAG for each distinct connected component $\overline{\mathbb{D}_j^m}$. This will the give us the updated set of running DAGs as $\overline{\mathbb{D}} = \overline{\mathbb{D}} \setminus \overline{\mathbb{D}_i} \cup \{ \overline{\mathbb{D}_j^m}  \mid \forall m \}$. 
Finally, we also remove the dataflow $D_r$ from the list of submitted DAGs, $\mathbb{D} = \mathbb{D} \setminus D_r$.

\subsection{Implementation}
        {\ysnoted{Architecture for Reuse Algorithms in Storm}}
We develop a \emph{Reusable Dataflow Manager} that offers a generic implementation of the merge and unmerge algorithms proposed above,  with bindings to an external DSPS to enact the dataflows and coordinate their reuse. We implement bindings for \emph{Apache Storm} DSPS due to its popularity for streaming applications, and to leverage existing IoT dataflow applications developed on it. Storm tasks are implemented in Java by extending a \texttt{Bolt} class, and a dataflow is called a \texttt{Topology} that wires the bolts together.

Users submit a dataflow to the manager as a JSON file which captures the tasks and their connectivity, including the task ID, type, and config. The manger keeps track of the state of the submitted and running DAGs. On running the merge algorithm, the manager identifies running DAGs that need to be merged, and new tasks and streams to be instantiated. 

DSPS like Storm do not allow the structure of a dataflow to be modified after launch, instead requiring it be stopped and a new dataflow with the updated DAG launched. This will be disruptive to all submitted dataflows that are supported by a running DAG. IoT domains are typically latency sensitive and may also have persistent mission-critical applications. As a result, we develop a mechanism to run the merged dataflows as partial DAGs that can be incrementally launched, and use a \emph{publish-subscribe broker} for externally connecting them. This is similar to an Enterprise Service Bus (ESB) model.

When the manager identifies multiple running dataflows to be merged ($\mathbb{Y}$) and new non-overlapping tasks and streams to be created ($\mathbb{T}_x,\mathbb{S}_{x}$), it takes the following steps. It first launches a new dataflow corresponding to the non-overlapping tasks and their local streams, $\langle \mathbb{T}_x,\mathbb{S}_{x}^{*} \rangle$. We use Storm's \emph{Flux} interface based on JSON to create and launch dataflows.  It then connects the boundary streams $\mathbb{S}_{x}^{*}$ from tasks in prior dataflows to tasks in the new dataflow through the broker. 

To this end, we have each task (\texttt{Bolt}) in the Storm dataflow extend our wrapper class, which subscribes to a unique \emph{control topic} on the broker to receive signals. The manager uses this topic to notify a task in a prior dataflow to \emph{forward} a copy of its output stream to a unique \emph{data topic}, which is subscribed to by the task in the new dataflow. Thus, the topic is a derived stream to connect tasks in different dataflows being merged. 

Similarly, when a submitted dataflow is to be demerged, the algorithm first identifies tasks and streams to terminate, $\langle \mathbb{T}_t,\mathbb{S}_{t} \rangle$. This would be in a single merged DAG per our algorithm. However, in our Storm implementation, a single merged dataflow may be deployed as multiple DAGs connected by the broker, with the manager doing the book-keeping. The tasks to be stopped may require termination of a subset of a running DAG, which Storm disallows. Instead, the manager notifies the tasks to be terminated to rather \emph{pause} their execution, using the control topic. These tasks will subsequently not process incoming messages. This, in effect, frees up resources of those tasks without disrupting the DAGs they belong to.

This approach can introduce \emph{latency overheads} due to the indirection in forwarding events between tasks through the broker. But it does not limit scaling since modern brokers like Kafka are designed for distributed scaling. Merging and demerging can also cause \emph{fragmentation} with many small Storm DAGs, though there may be fewer logical merged DAGs, and several paused tasks. In future, we can perform periodic \emph{defragmentation}, where running Storm DAGs are stopped, and a single DAG relaunched for each merged dataflow.

\section{Experiments}
\label{sec:results}
\begin{figure*}[t]
	\centering	
	\subfloat[OPMW, Sequential]{%
		\includegraphics[width=0.35\textwidth]{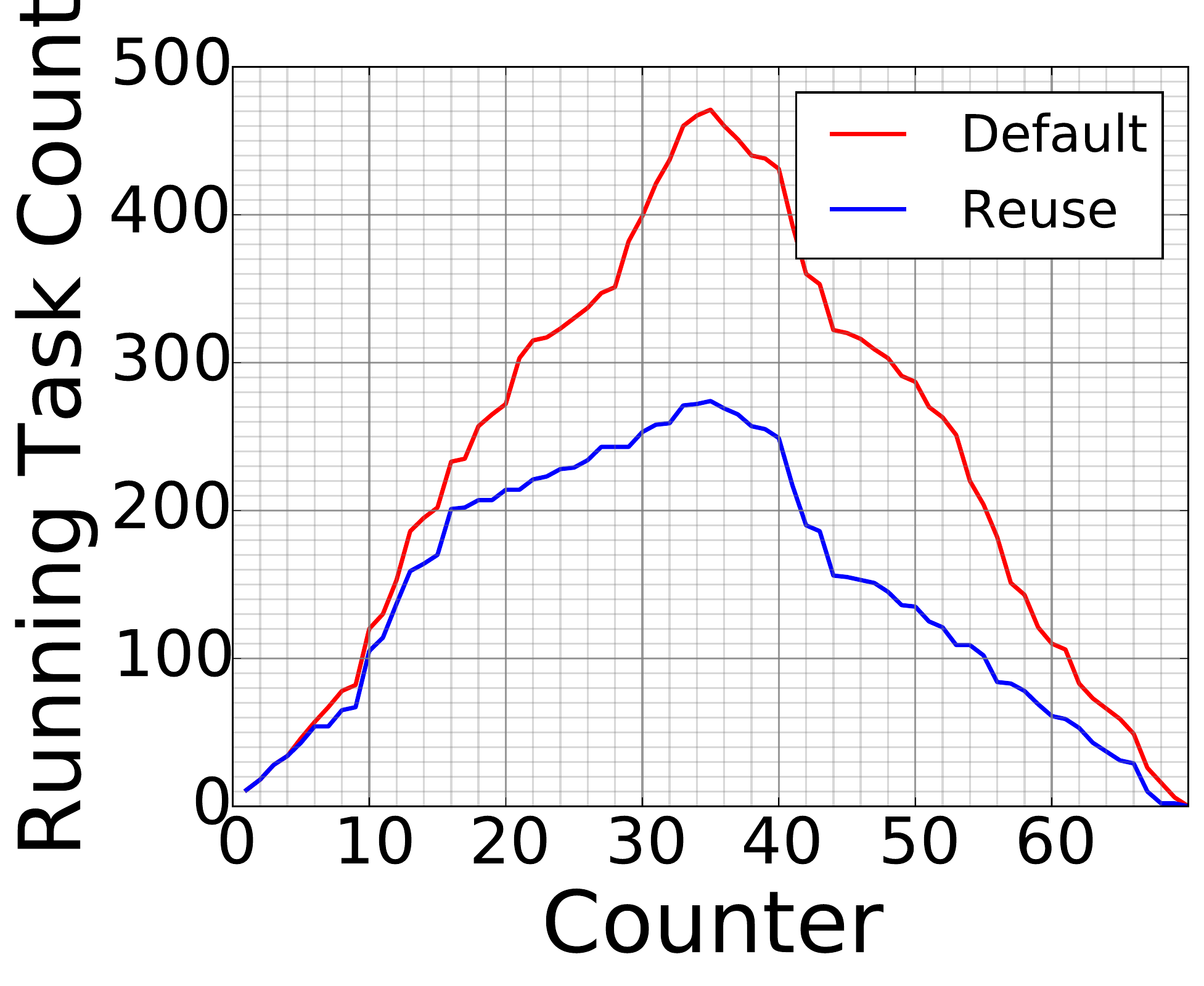}
                \label{fig:opmw:seq:tasks}
	}
	\subfloat[OPMW, Rnd Walk 1 ]{%
		\includegraphics[width=0.35\textwidth]{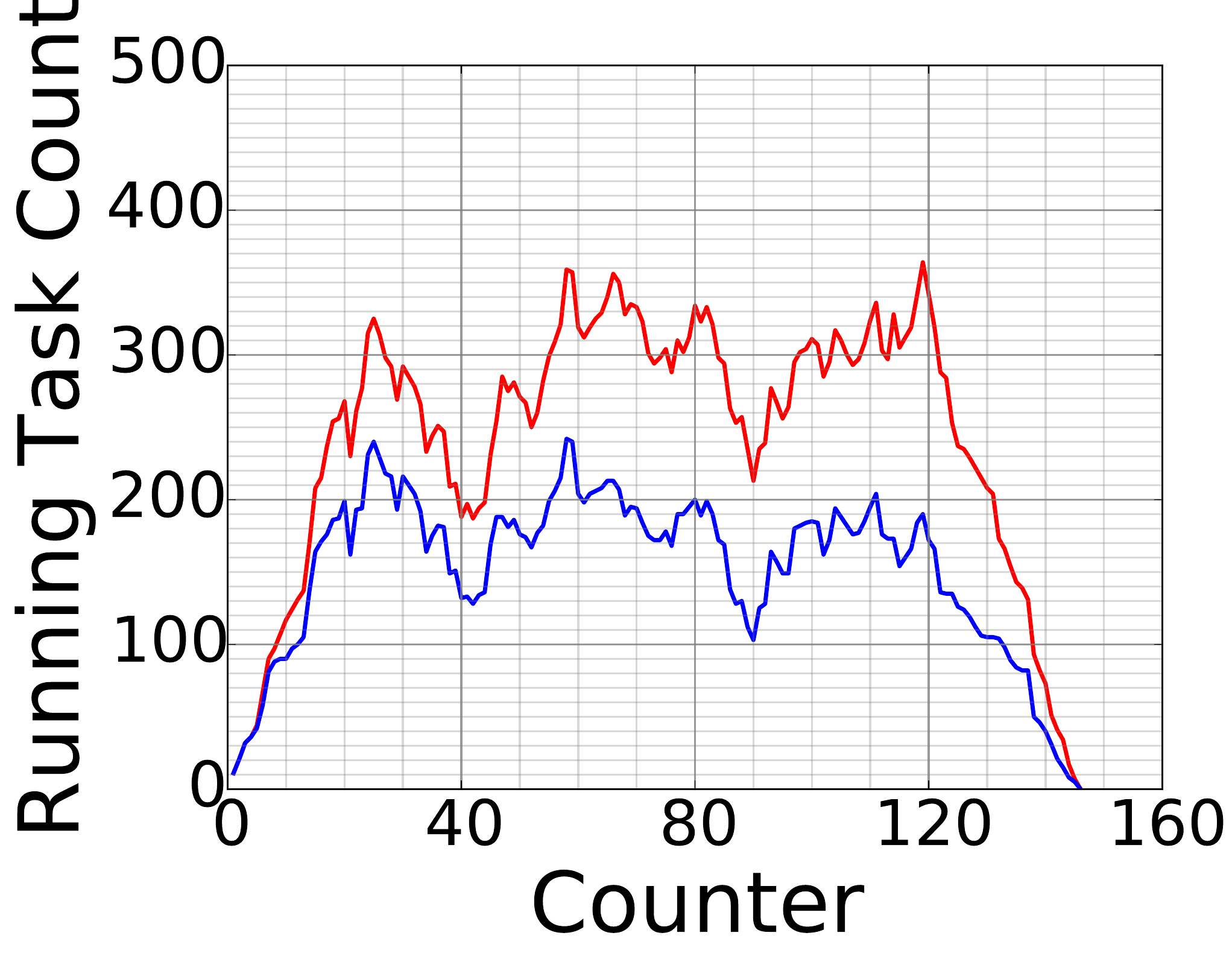}
		\label{fig:opmw:rw1:tasks}
	}
	\subfloat[OPMW, Rnd Walk 2 ]{%
		\includegraphics[width=0.35\textwidth]{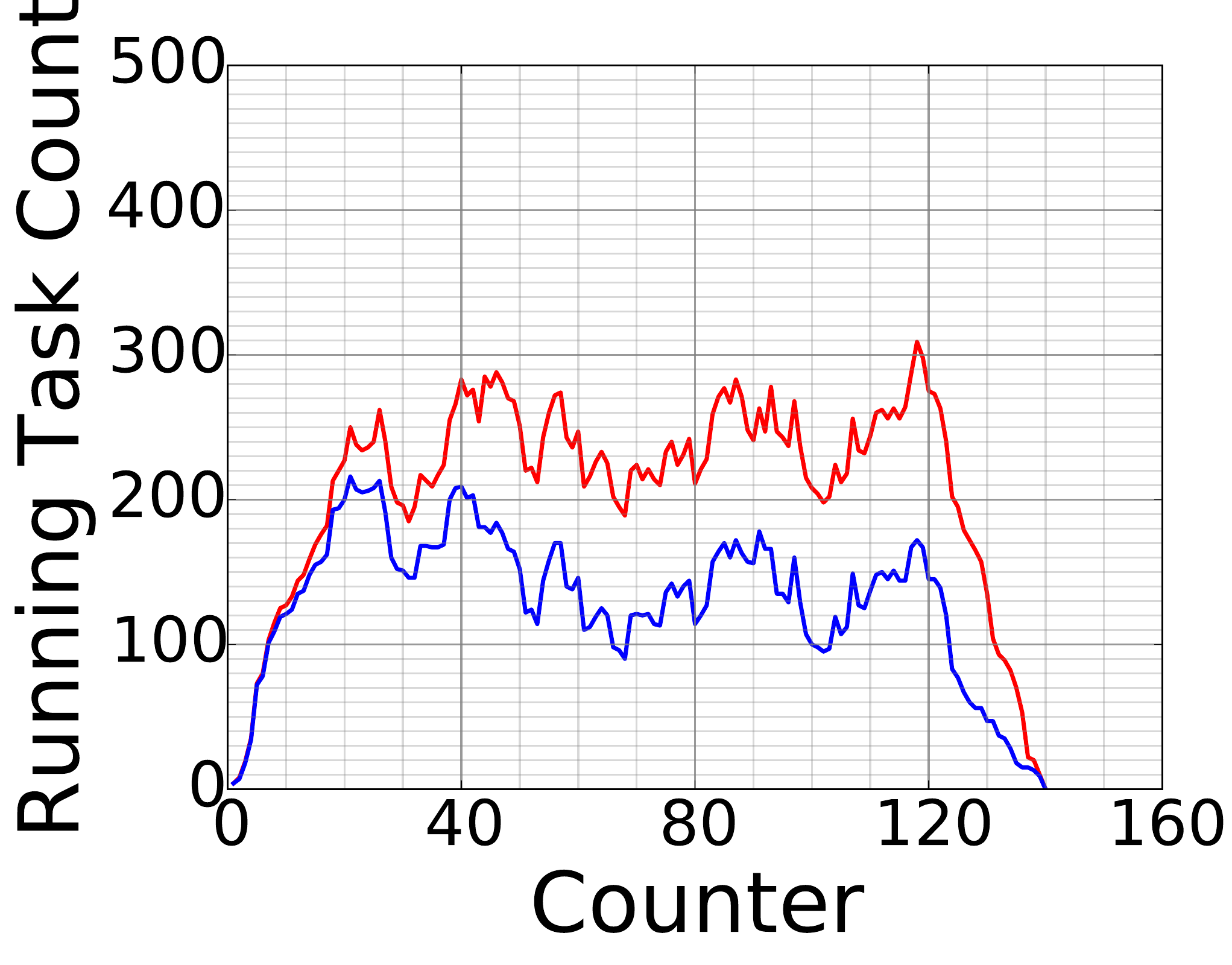}
		\label{fig:opmw:rw2:tasks}
	} \\
        \subfloat[RIoT, Sequential]{%
		\includegraphics[width=0.35\textwidth]{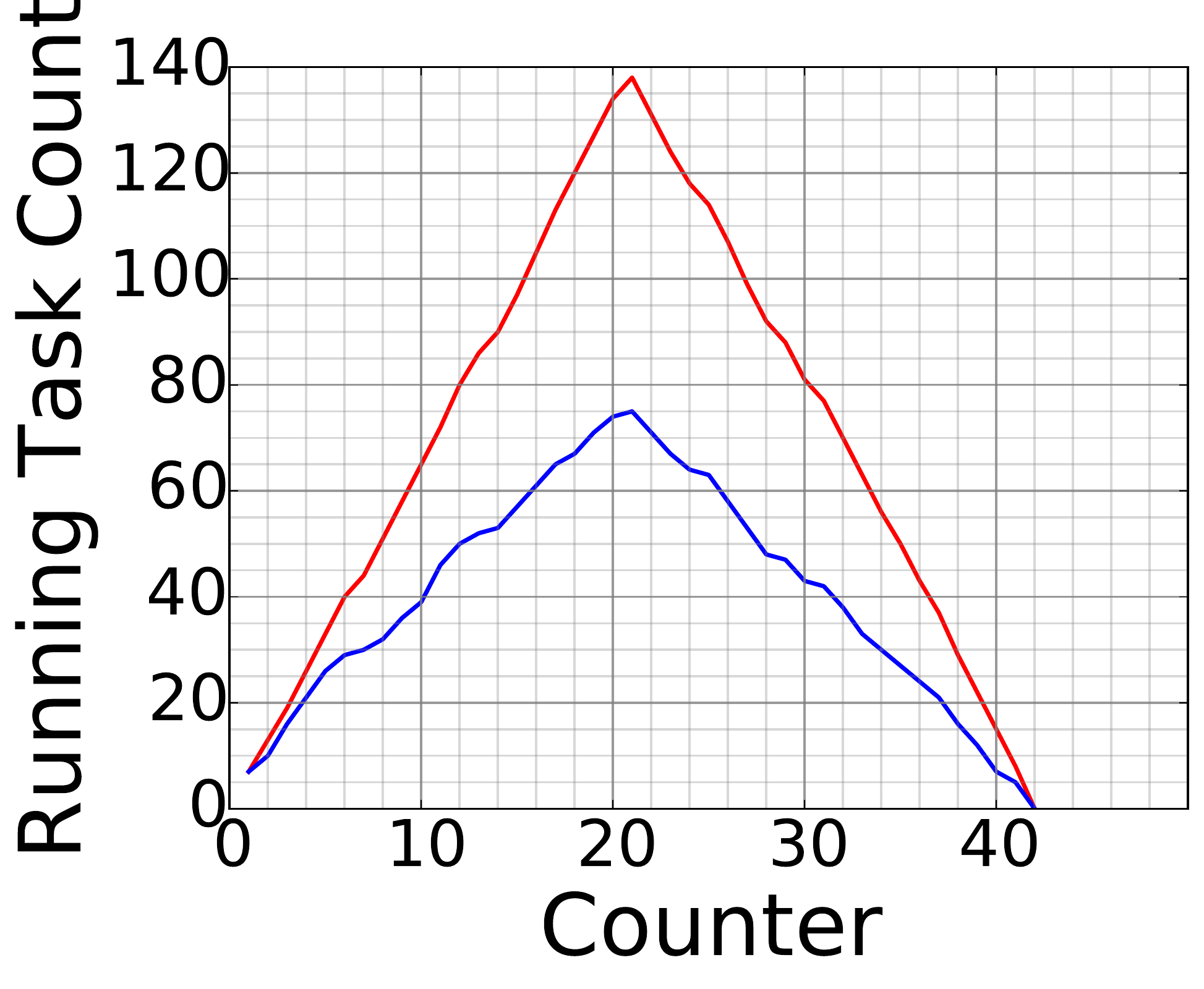}
		\label{fig:riot:seq:tasks}
	}
	\subfloat[RIoT, Rnd Walk 1]{%
		\includegraphics[width=0.35\textwidth]{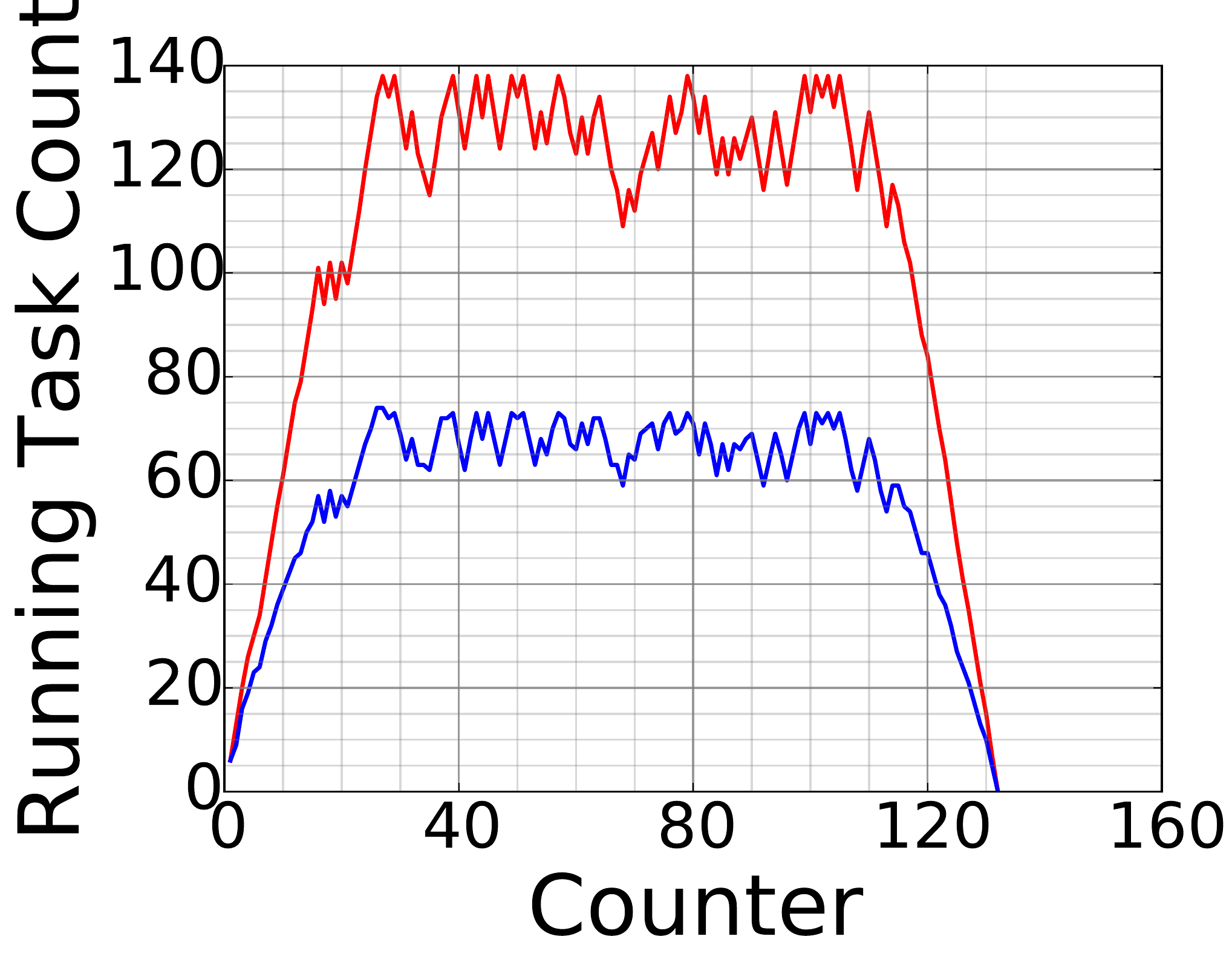}
		\label{fig:riot:rw1:tasks}
	}
	\subfloat[RIoT, Rnd Walk 2]{%
		\includegraphics[width=0.35\textwidth]{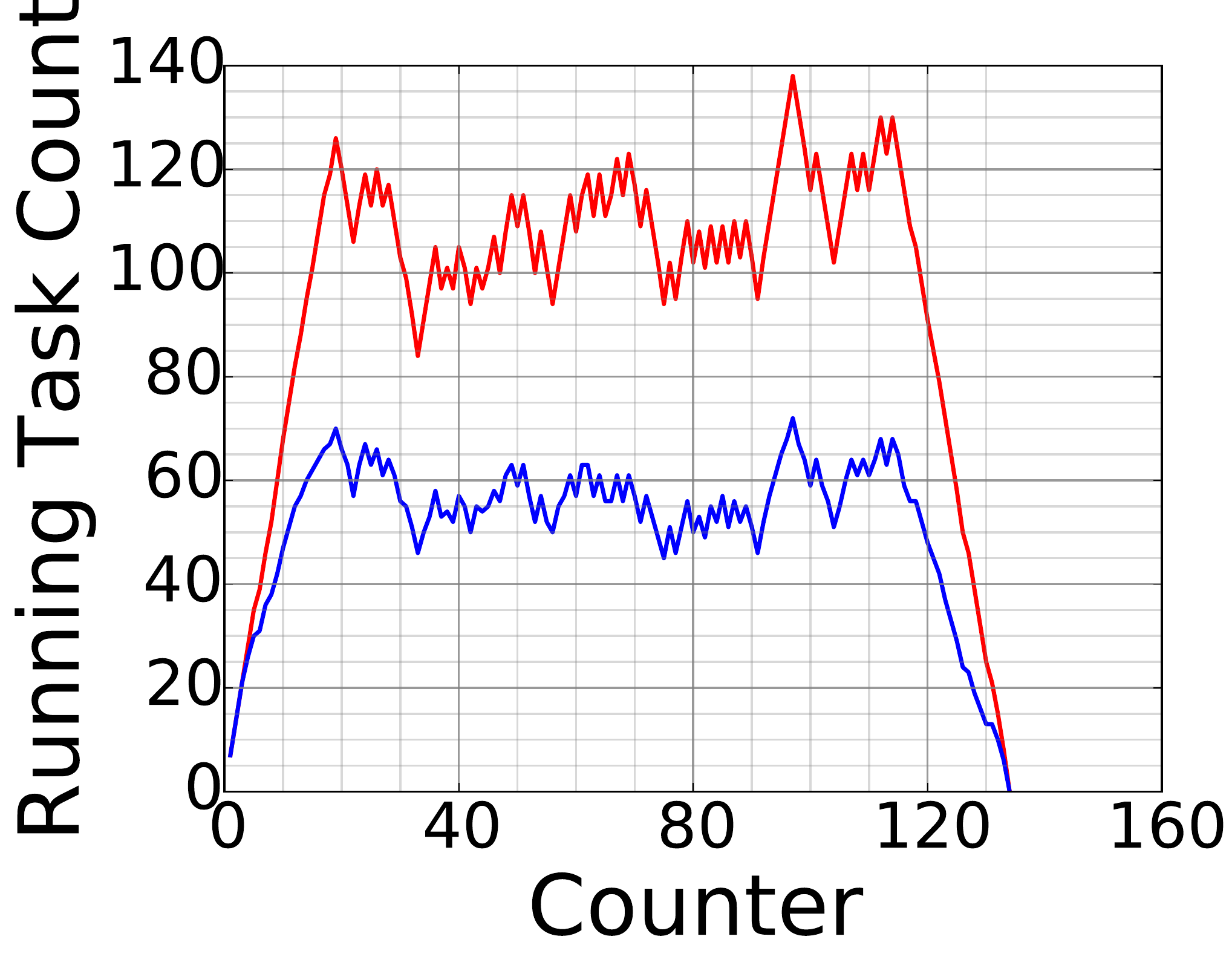}
		\label{fig:riot:rw2:tasks}
	}
	\caption{\emph{Number of running tasks} at different time points for the 6 workloads, using Storm Default and with Reuse.}
	\label{fig:tasks}
\vspace{-0.17in}
\end{figure*}
\begin{figure*}[t!]
	\centering
	\subfloat[OPMW, Sequential]{%
		\includegraphics[width=0.35\textwidth]{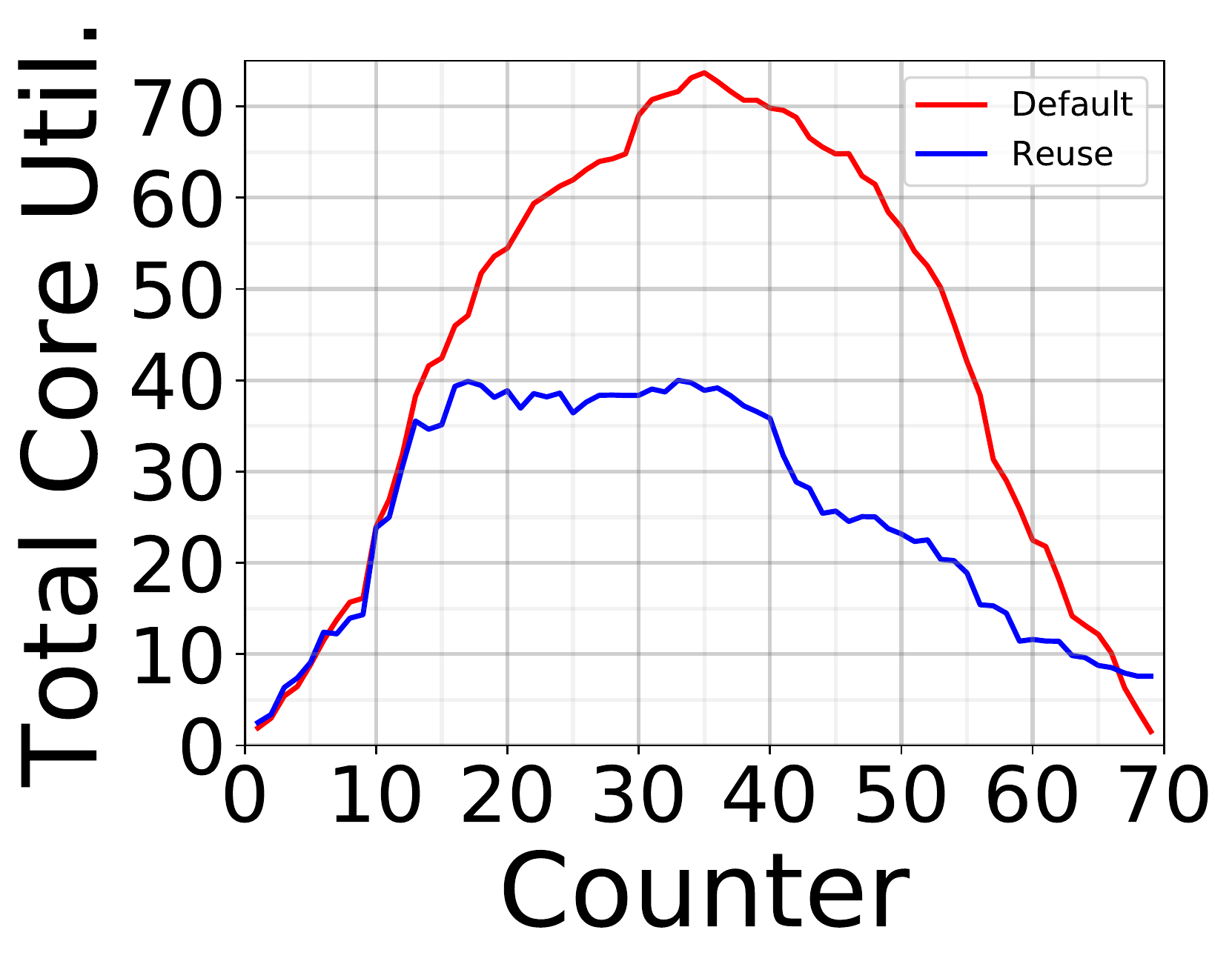}
                \label{fig:opmw:seq:cpu}
	}
	\subfloat[OPMW, Rnd Walk 1]{%
		\includegraphics[width=0.35\textwidth]{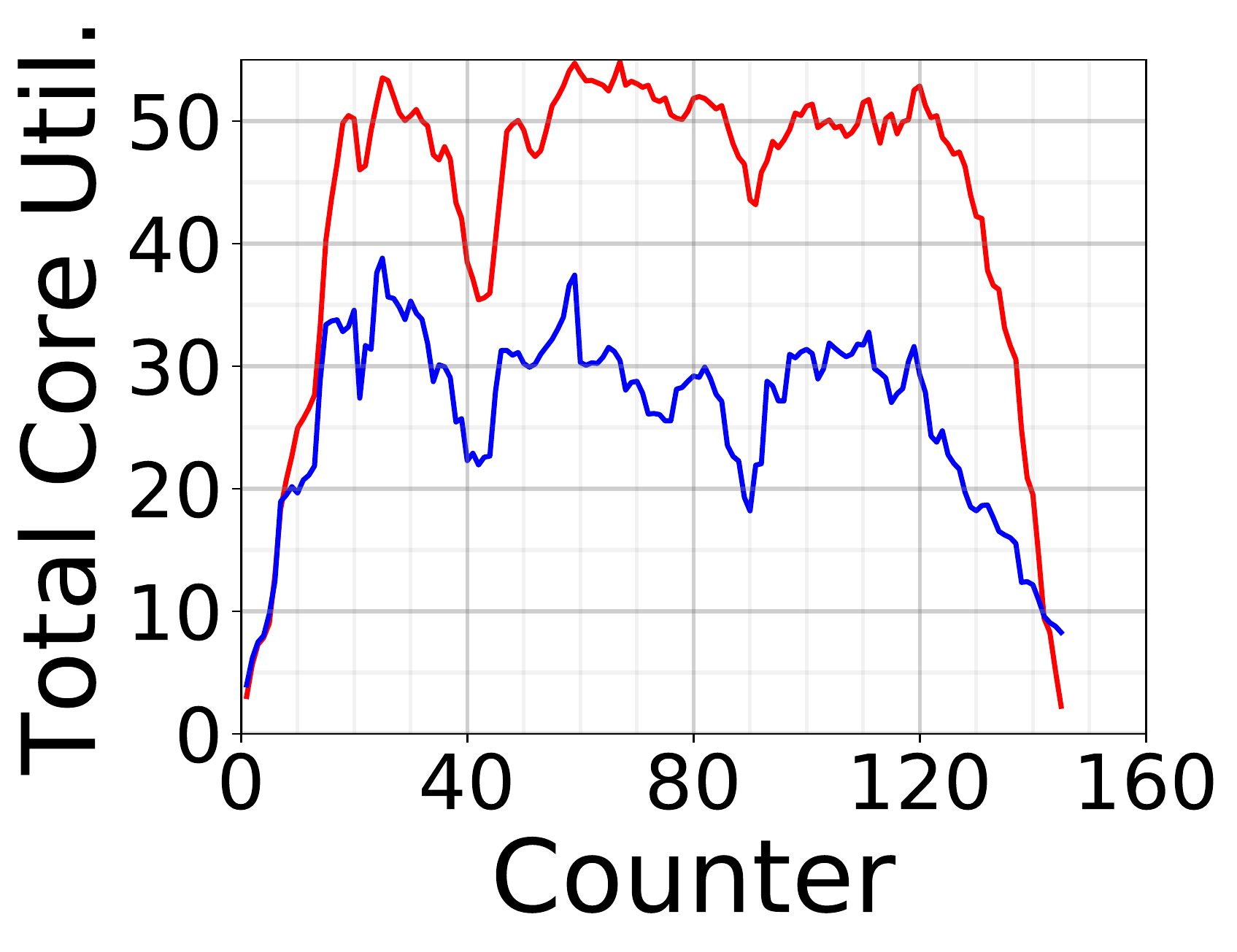}
		\label{fig:opmw:rw1:cpu}
	}
	\subfloat[OPMW, Rnd Walk 2]{%
		\includegraphics[width=0.35\textwidth]{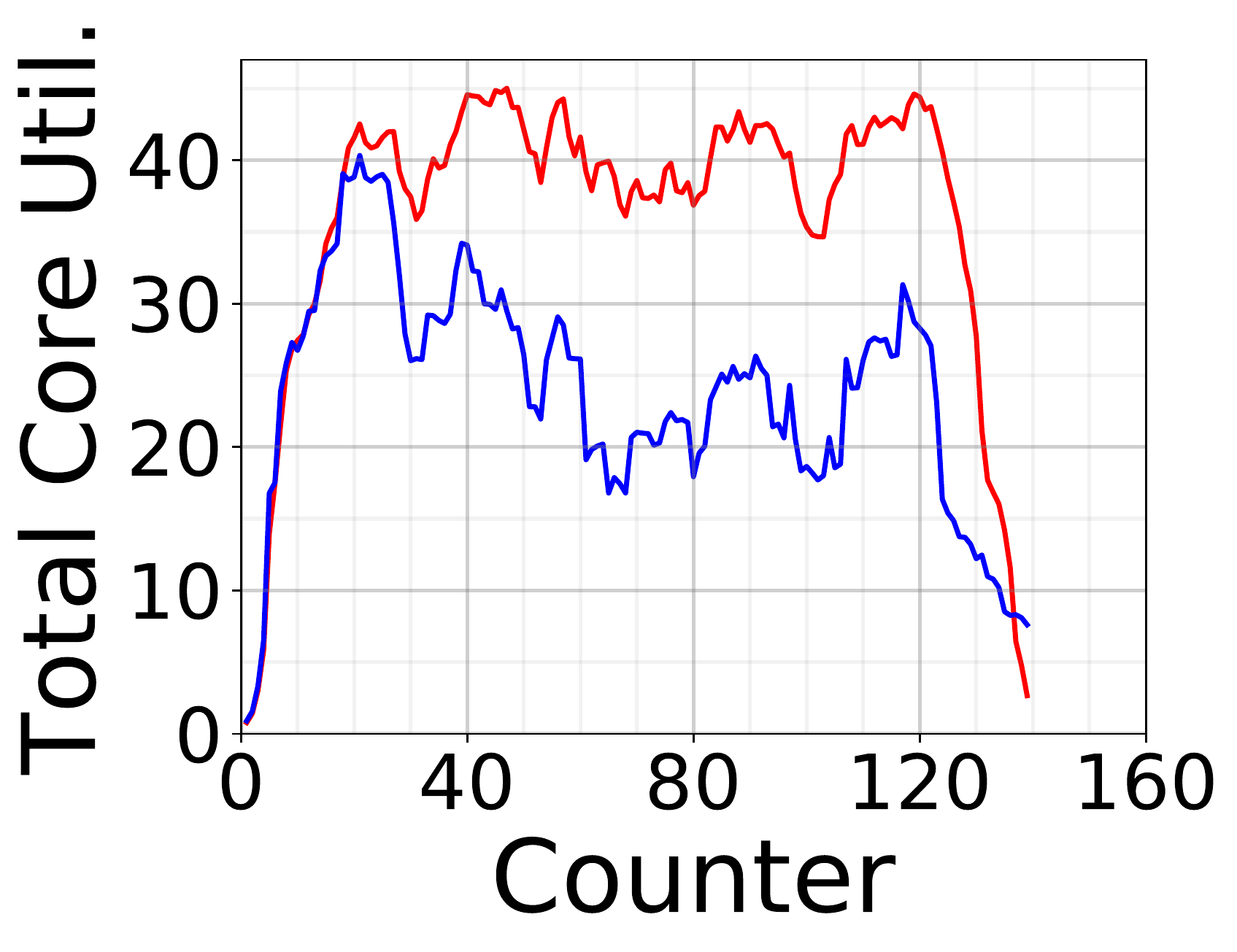}
		\label{fig:opmw:rw2:cpu}
	} \\
        \subfloat[RIoT, Sequential]{%
		\includegraphics[width=0.35\textwidth]{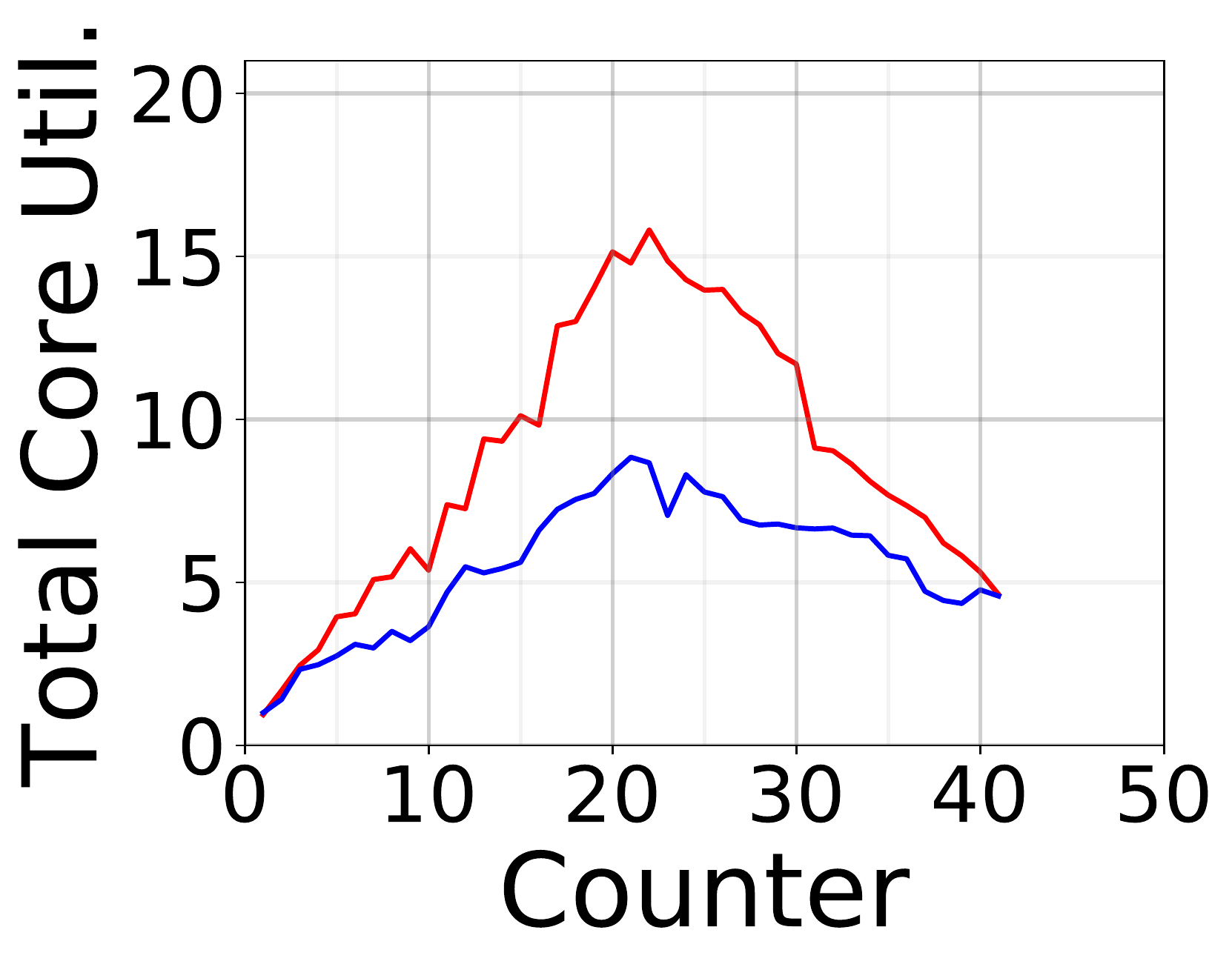}
		\label{fig:riot:seq:cpu}
	}
	\subfloat[RIoT, Rnd Walk 1]{%
		\includegraphics[width=0.35\textwidth]{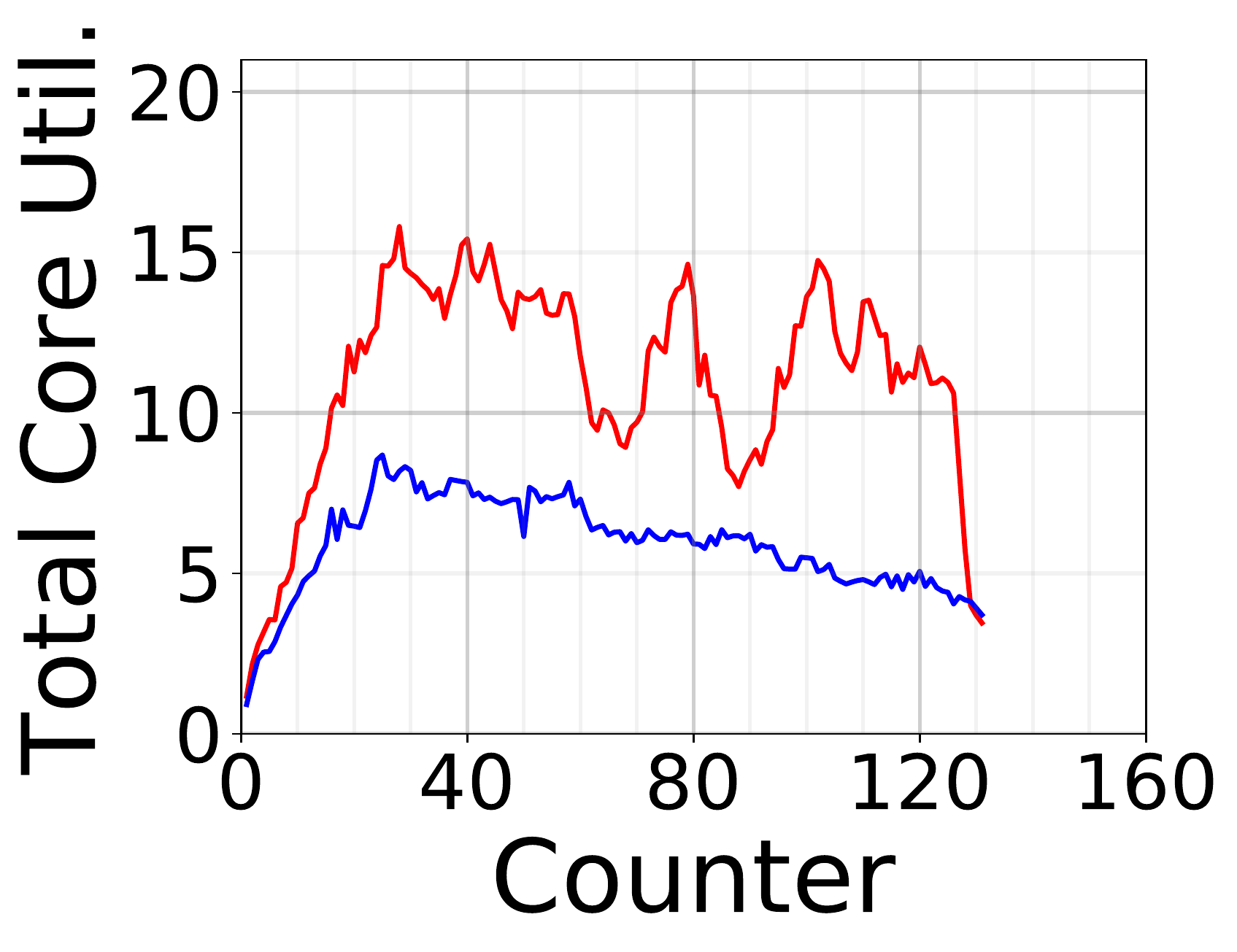}
		\label{fig:riot:rw1:cpu}
	}
	\subfloat[RIoT, Rnd Walk 2]{%
		\includegraphics[width=0.35\textwidth]{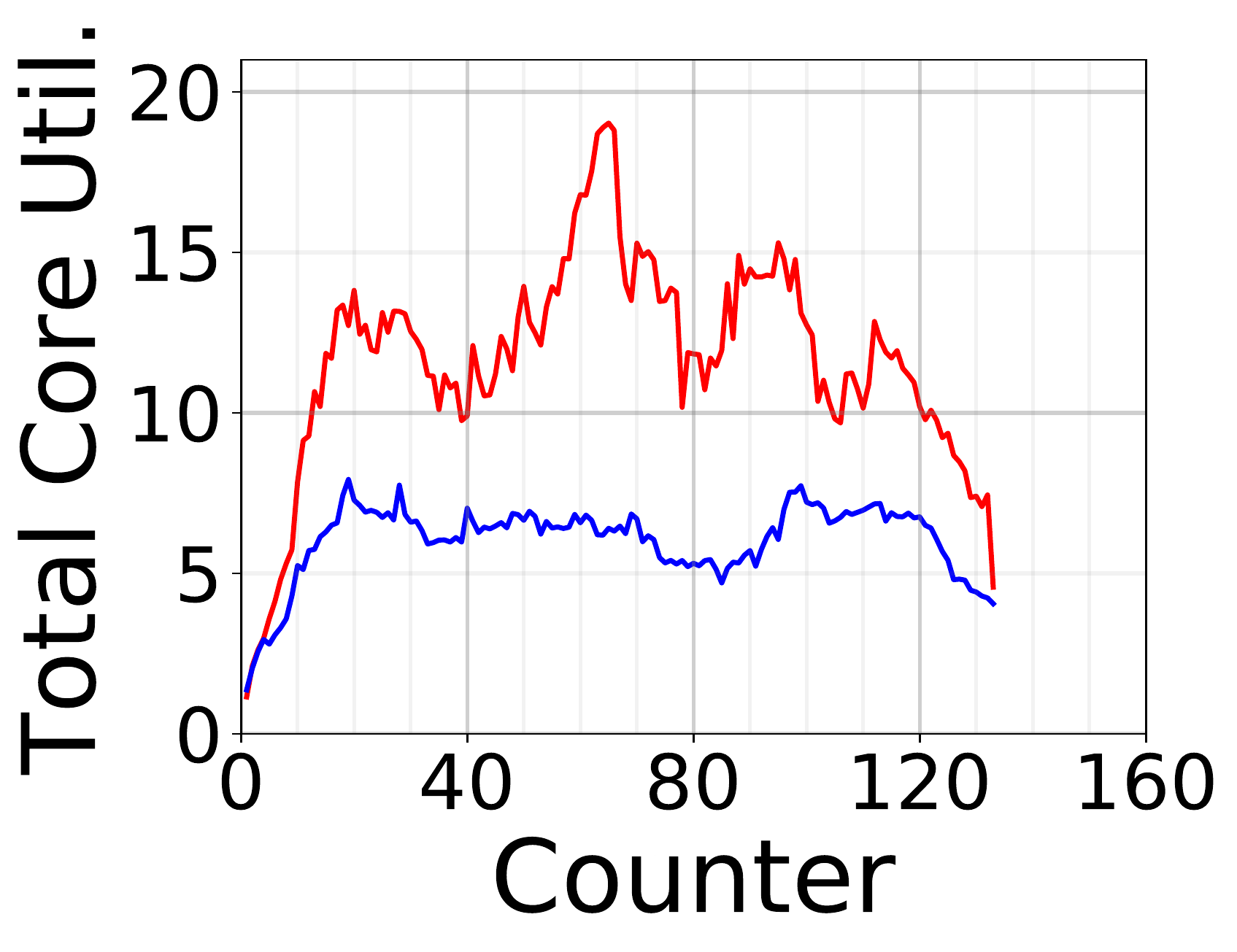}
		\label{fig:riot:rw2:cpu}
	}
	\caption{\emph{Cumulative Core usage} at different time points for the 6 workloads, using Storm Default and with Reuse.}
	\label{fig:cpu}
\vspace{-0.17in}
\end{figure*}
\begin{figure*}[t!]
	\centering
	\subfloat[OPMW, Sequential]{%
		\includegraphics[width=0.35\textwidth]{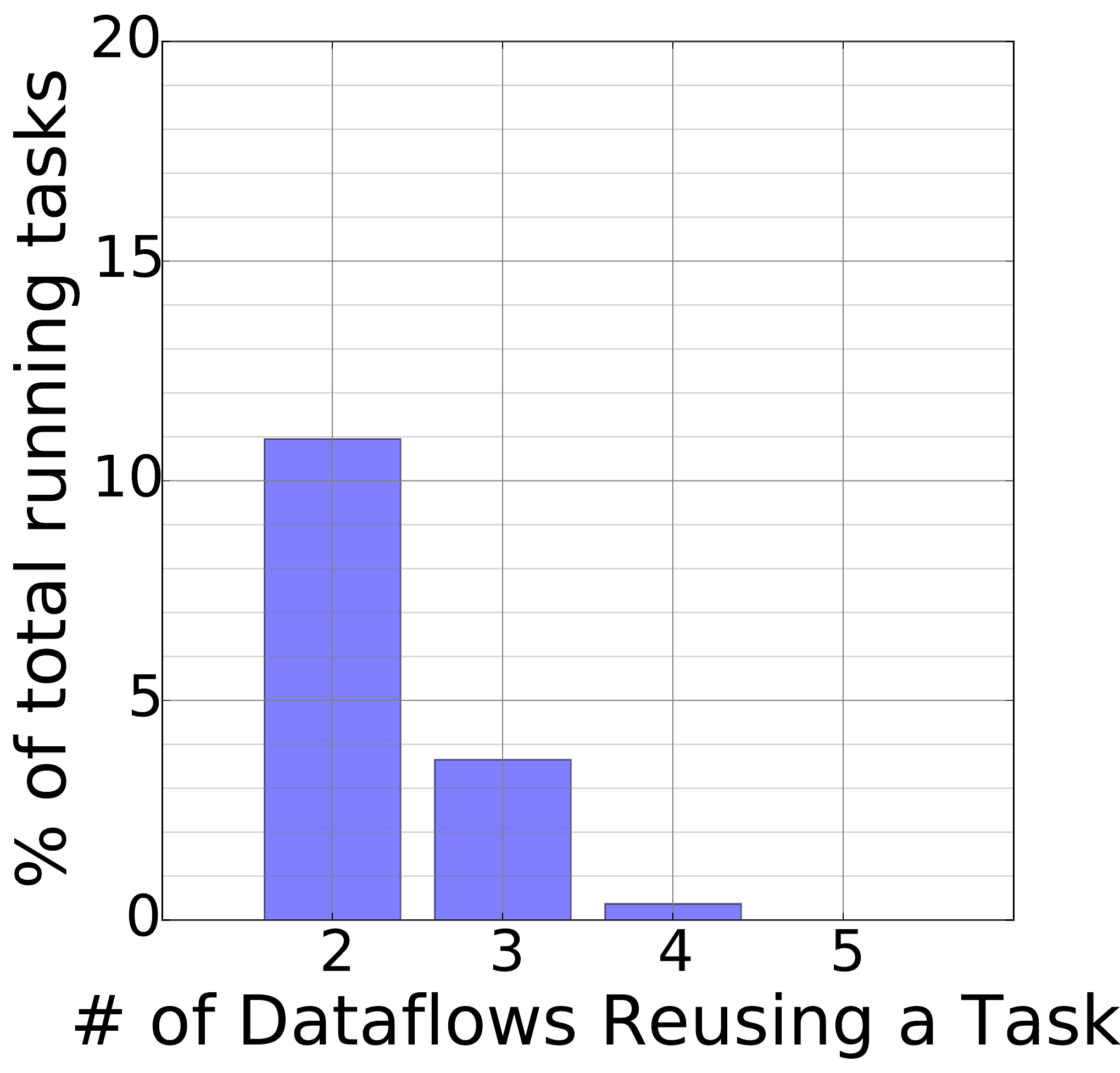}
                \label{fig:opmw:seq:freq}
	}
	\subfloat[OPMW, Rnd Walk 1]{%
		\includegraphics[width=0.35\textwidth]{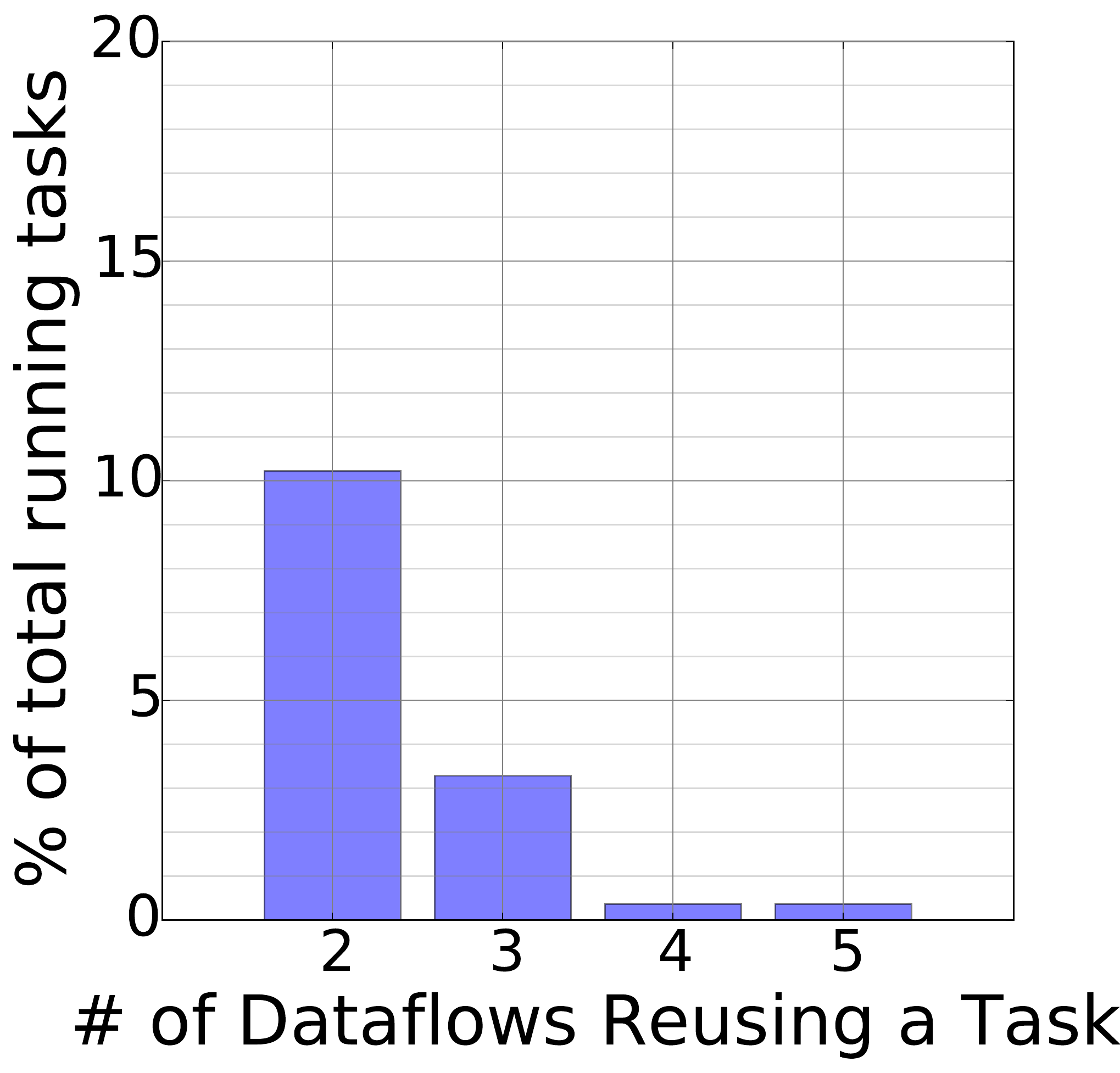}
		\label{fig:opmw:rw1:freq}
	}
	\subfloat[OPMW, Rnd Walk 2]{%
		\includegraphics[width=0.35\textwidth]{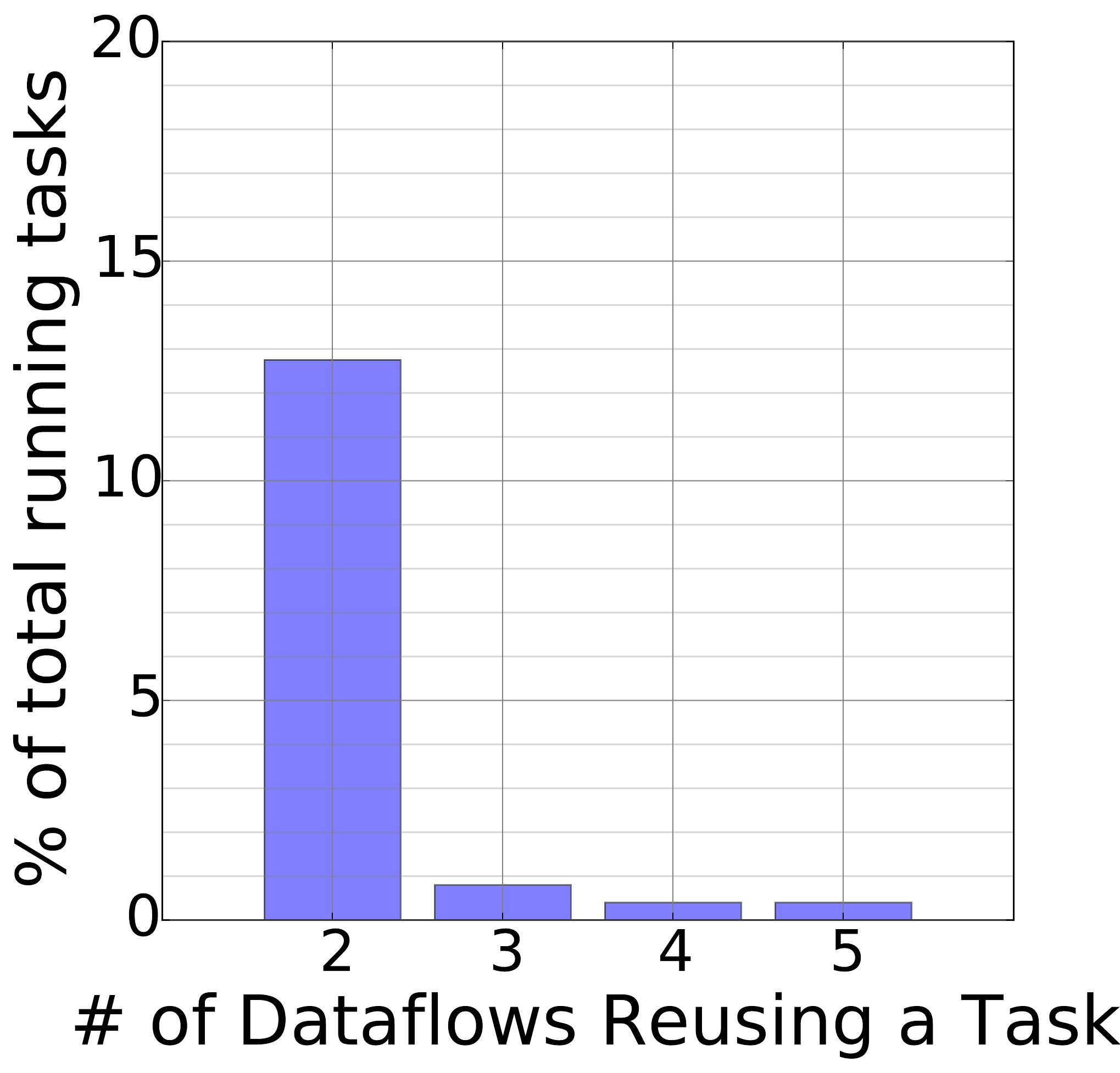}
		\label{fig:opmw:rw2:freq}
	} \\
        \subfloat[RIoT, Sequential]{%
		\includegraphics[width=0.35\textwidth]{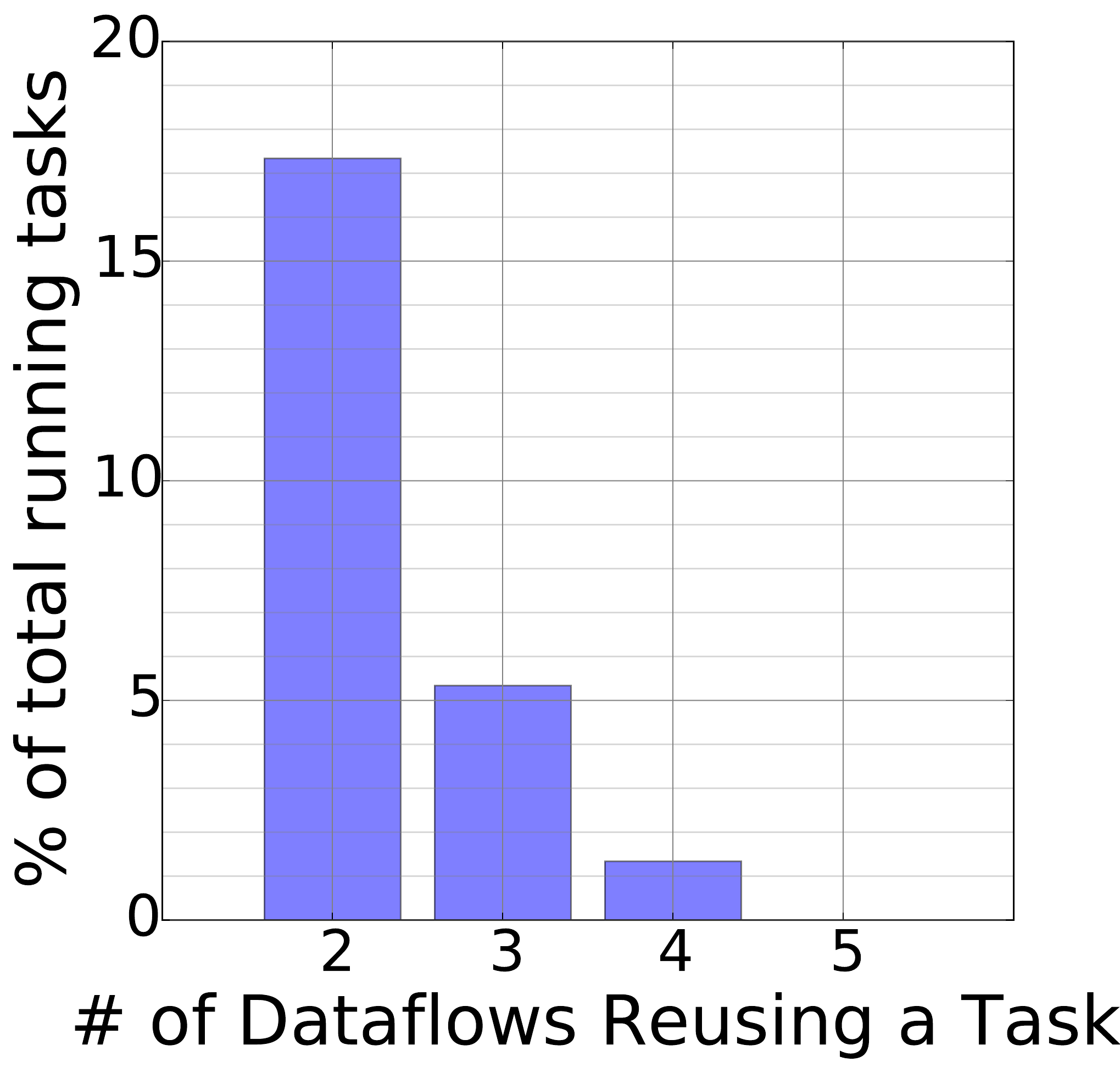}
		\label{fig:riot:seq:freq}
	}
	\subfloat[RIoT, Rnd Walk 1]{%
		\includegraphics[width=0.35\textwidth]{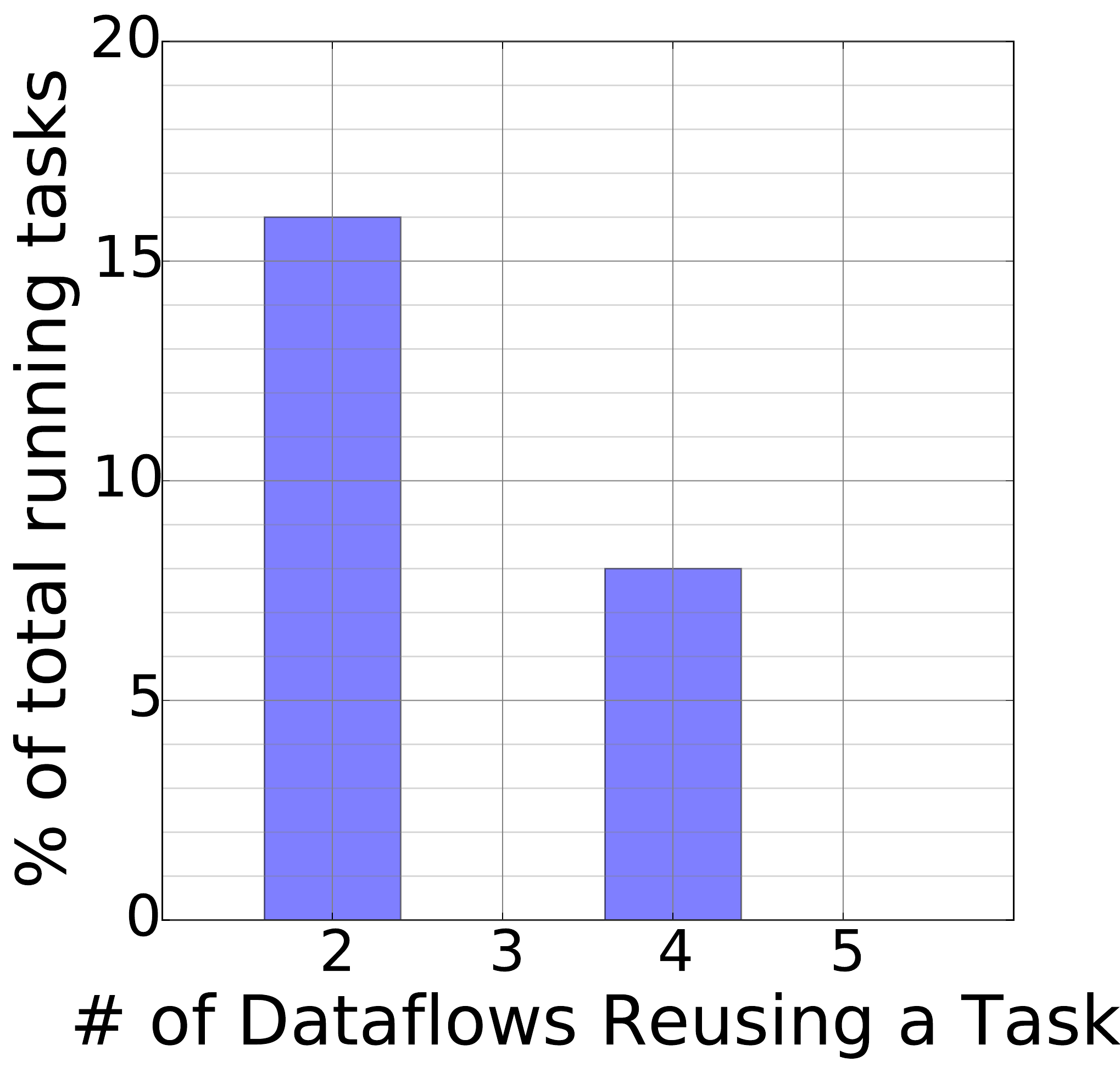}
		\label{fig:riot:rw1:freq}
	}
	\subfloat[RIoT, Rnd Walk 2]{%
		\includegraphics[width=0.35\textwidth]{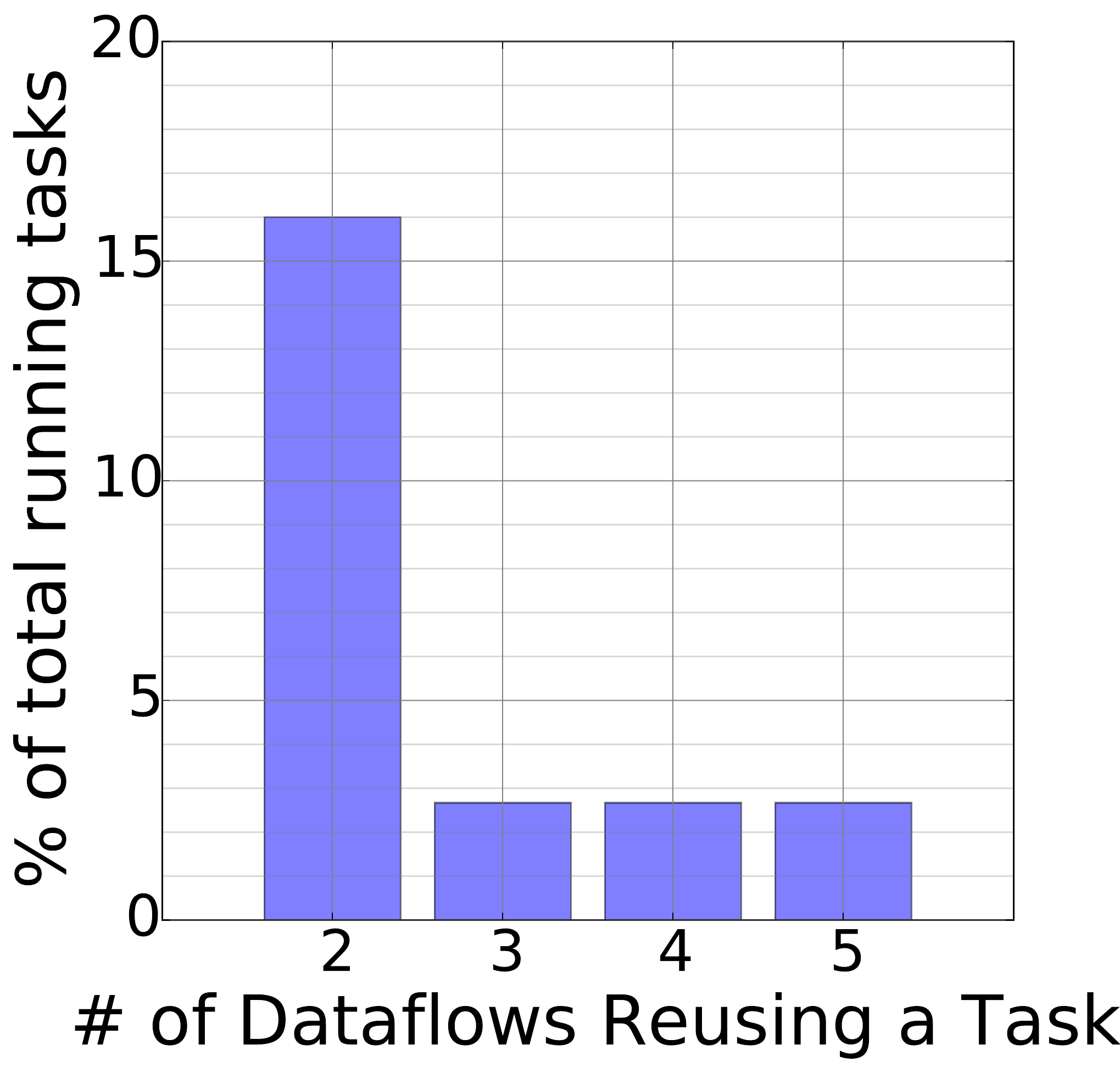}
		\label{fig:riot:rw2:freq}
	}
	\caption{\emph{Frequency of reuse} histogram showing the fraction of time that tasks were (re)used by more than one dataflow.}
	\label{fig:freq}
\vspace{-0.17in}
\end{figure*}

\subsection{Workloads}
We use two workloads in our evaluation. One is from the \emph{Open Provenance Models for Workflows ({OPMW})} repository~\cite{opmw} which hosts ontology-based scientific workflow  models and their traces. These workflows span different domains, and are designed to be shared by the science community. As there are few publicly available streaming IoT dataflows, these OPMW workflows are a proxy for future IoT dataflow collections in the public domain. 

Of the 74 usable OPMW workflows available in the portal, we choose 35 arbitrary ones such that they can cumulatively fit in our cluster. 
These have $471$ total tasks, of which $219$ are unique, with $2-38$ tasks present in each DAG. We only use the OPMW DAG structure, with the task ID, type, and their connectivity. In the Storm implementation of the DAGs, we replace the internal task logic with an iterative computation of $\pi$ that outputs a floating point number. This reduces dependencies while ensuring that each input event triggers a CPU-intensive operation.


The second workload is based on real IoT applications that are available as part of our \emph{Real-time IoT Benchmark (RIoTBench)}~\cite{riotbench}. The benchmark has over 30 stream processing tasks for IoT domains, classified as parsing and filtering, I/O, statistical and visual analytics, etc., These are composed into $4$ streaming IoT dataflows for Extract Transform Load (ETL), Statistical Summarization, and Predictive Analytics. We extend these dataflows with additional permutations of their DAGs from the available tasks and use $3$ IoT source tasks -- Smart Power Grid, Urban Sensing, and  NY City Taxi Cab streams -- to construct $21$ different IoT applications with real logic. These DAGs have $138$ total tasks with $19$ distinct ones.


We generate $3$ DAG traces each for the OPMW and RIoT dataflows to simulate submission and removal. For one trace, we use a \emph{Sequential Submit/Drain {(SEQ)}} model to first incrementally \emph{submit} a random dataflow from the workload with uniform probability, without replacement, in each time step. Once all DAGs in the workload are added, we switch to a \emph{drain} phase where a random DAG that was submitted and present is removed in each step. This takes $35 \times 2 = 70$ time steps for OPMW and $21 \times 2 = 42$ for RIoT. This trace simulates the behavior when only add or only remove operations occur, and the maximum reuse happens when all DAGs are submitted. 

For the two other traces, we generate \emph{Random Walks {(RW)}} where we perform an add or a remove with equal probability at each time step, and repeat this $100$ times. The DAGs to add/remove are chosen at random from the available/submitted pool -- a submitted DAG is not resubmitted (unless removed) to avoid the whole DAG being reused by our algorithm to unfairly inflate its benefits. We initially populate the system with $20$ DAGs for OPMW and $15$ DAGs for RIoTBench at random, which are $\approx\frac{2}{3}^{rds}$ of the workload before the random walk starts, and similarly drain the system after the $100$ random walks. 
These traces help evaluate the impact on the system after repeated merge and demerge operations, and also test for any inconsistencies under sustained operations.



\subsection{Setup}
We run our experiments on \emph{Apache Storm v1.0.2} DSPS that is setup on a commodity cluster, with each node having an AMD Opteron $3380$ 8-core CPU@$2.6$~GHz, $32$~GB RAM, a $256$~GB SSD, and GigaBit Ethernet, running CentOS v7. Storm runs on JRE v1.8 with the \emph{Flux} JSON interface used for dataflow submission. \emph{Apache Apollo v1.7.1} is our publish-subscribe broker using the MQTT protocol. Our \emph{Manager} is implemented in Java and talks to Storm from a local node.

DAGs submitted to Storm use the default parallelism of 1~thread per task. Storm uses a round-robin scheduler to assign tasks to \emph{Workers} in its cluster. Each node runs the default one Worker JVM per core, and we see that up to 8 tasks can run on a Worker without interference. This means up to $64$ tasks can be placed on a single node. However, each Worker can have tasks from only one DAG. The Storm cluster is assigned as many nodes as required at the peak of a given trace, and this ranges from $3-9$ machines depending on the trace.

Each \emph{action} in a trace -- submission or removal of a DAG -- is sent directly to the Storm service for the \emph{Default} scenario without reuse, or sent to our Manager when using the \emph{Reuse} algorithms. The actions are generated at a fixed time step of $1~min$ to allow the system and resource usage to stabilize. So each run of a trace takes between $42-140~mins$ for the Default and Reuse approaches. At each time step, we report the total running task count (Figs.~\ref{fig:tasks}) and use \texttt{top} to measure the cumulative CPU utilization across all nodes; this is reported as the number of cores used in Figs.~\ref{fig:cpu}, where 100\% cumulative CPU in the cluster $\Rightarrow$ 1~core used. We used a constant input rate of $10~events/sec$ for each source task as this matches the peak rate supported for the given resource allocation. Each event is $4-380~bytes$ in size, depending on the source.

\subsection{Results for OPMW Workload}
In the \emph{sequential workload}, we do not initially see a big gap in the task counts between Default and Reuse, until step $16$ (Fig.~\ref{fig:opmw:seq:tasks}). At this point, the Default strategy has $233$ tasks while our Reuse strategy has $201$ tasks running. The random selection of DAGs for addition happens to select dataflows with few equivalent tasks for reuse.

However, as more of the dataflows are added till the whole workload is deployed at step $35$, we see this gap widen, with Default running $471$ tasks while Reuse has only $274$ active ones. This stark contrast at the peak highlights the maximum possible reuse being exploited when all available dataflows are submitted.
In the drain phase, since the DAGs are also removed in random order, the gap stays wide since sampling happened to remove DAGs with less reuse first.


Fig.~\ref{fig:opmw:seq:cpu} shows the cumulative sum of the CPU core usage across all active hosts on the Y axis as DAGs are added and removed; a value of 1 implies $100\%$ use of 1 core. We see a strong correlation between the number of running tasks and the cores used. Until step 12, 
they both consume about the same number of cores, reaching $\approx31$. Beyond that, the usage plateaus out for Reuse at $\approx38$ cores as tasks get reused and task-count flattens, while it continues to grow for Default to reach a peak of $74$ cores with $35$ DAGs. At the peak, Reuse takes $42\%$ fewer CPU resources, which directly translates into monetary savings on public Cloud VMs.
We do notice that when the dataflows are drained, the core usage reduces for both approaches, though at different slopes. Interestingly, there is a cross-over at step $67$ when usage for Reuse is higher at $8$ while Default takes $6$, even though Reuse runs fewer tasks. This is due to the overhead of \emph{pause}. Even when a task is paused in the Reuse approach, it continues to consume minimal resources as it is still deployed within Storm. This accumulates with increased DAG fragmentation, and toward the end, all $274$ tasks that were once running but now in a paused state end up consuming $7.5$ cores. This motivates the need for periodic \emph{defragmentation}, when all DAGs are stopped and a single Storm dataflow started for each merged DAG.

Lastly, Fig.~\ref{fig:opmw:seq:freq} shows the histogram of the time-weighted fraction of all running tasks over all steps (Y axis) that were reused by $[1,\underline{2})$ DAGs, $[2,\underline{3})$ DAGs, etc. (X axis). We omit the frequency of tasks used just once, that is the residual of all these frequencies.  We see that $11\%$ of all tasks are reused by $> 1$ and $\le2$ dataflows, while another $4\%$ are reused by $>2$ dataflows. But even this small fraction of reuse is helping achieve significant reduction in resource needs. 

The \emph{Random Walk} workload traces (RW1 and RW2) have $\approx20$ median number of dataflows that are active. RW1 has a wider oscillation of the task counts than RW2 (Figs.~\ref{fig:opmw:rw1:tasks} and~\ref{fig:opmw:rw2:tasks}) due to more contiguous adds/removes of dataflows. For RW1, the Default's task count ranges from $187-364$ during the $100$ walks while Reuse has $103-242$ tasks in this period. For RW2, these ranges are respectively $185-309$ and $90-216$. While the task count for Reuse parallels the Default, it always maintains an advantage for all $100$ steps, and on an average has $38\%$ fewer running tasks. The reuse histograms (Fig.~\ref{fig:opmw:rw1:freq},~Fig.~\ref{fig:opmw:rw2:freq}) indicate that $14\%$ of the tasks were used by more than one DAG.

This has a corresponding impact on the cores used by these traces, where Default takes a median of $50$ and $41$ cores for RW1 and RW2, while Reuse takes only $30$ and $26$ -- a cost savings of $37-40\%$. 
While the impact of fragmentation in seen here as well, with $\approx 8$ cores remaining in use at the end, defragmentation can even help improve utilization under normal conditions if adequate tasks are paused. 

\ysnoted{Extension: do periodic defragmentation and measure improvement in util\%}
\ysnoted{Extension: Report makespan. Do periodic defragmentation and measure improvement in makespan}


\subsection{Results for RIoT Workload} 
%
%
We see that in the \emph{sequential trace} of the RIoT workload, the gap between Default and Reuse strategies starts right from step 2, with the running task counts growing smoothly in both cases but at different rates (Fig.~\ref{fig:riot:seq:tasks}). The RIoT DAGs are more homogeneous, with each DAG having $4-8$ tasks, compared to OPMW where the had $2-38$ tasks. 
At the peak submission in time step $21$, the running tasks count reaches $138$ with Default while it is only $75$ using the Reuse approach.

The correlation between the number of running tasks and the core usage is seen for this workload too. We observe an average reduction in cores used by $37.5\%$ 
(Fig.~\ref{fig:riot:seq:cpu}), which is more than for OPMW. Their reuse histogram also shows $24\%$ of tasks used by more than one dataflow (Fig.~\ref{fig:riot:seq:freq}). RIoT applications come from the same domain, and their potential for reuse is therefore higher. However, since they incorporate real IoT task logic rather than just a generic $\pi$ logic, the core usage for RIoT shows more variability since the DAG tasks have diverse computing needs. This is despite RIoT DAGs having a similar number of tasks. 
So while task counts and task reuse counts offer a quick approximation of resource benefits, the core usage gives a more accurate sense of cost savings. 

The benefits are even more evident for the \emph{random walk traces} of RIoT. Here again the task count is less variable across time, with a mean of $125 \pm 25\%$ and $67 \pm 22\%$ running tasks for Default and Reuse on RW1. However, the core usage varies more due to the IoT task logic, with a mean of $12 \pm 37\%$ for Default and $6 \pm 34\%$ for Reuse. Similar values are seen for RW2 as well. The reuse histograms also shows that $8\%$ of tasks were used by 2 or more dataflows. These all translate to an enhanced cost benefit for Reuse, with core usage $47-51\%$ lower than Default, for both the random walks.



In summary, we see that the Reuse strategy offers significant reduction in the running task count along with real cost savings of up to $51\%$ lower core usage, relative to the Default approach. These apply both to public dataflows from the OPMW multi-domain repository with synthetic task logic, and to permutations of real IoT dataflows with diverse task logic of RIoTBench. Further improvements also seem achievable for the Reuse approach if defragmentation is done as well.
\ysnoted{Run for all 74 OPMW workflows}
\ysnoted{Run for longer permutations of RioT workflows}

\section{Related Work}
\label{sec:related}
There are two broad categories of ``reuse'' research that are relevant to our problem: \emph{distributed stream processing applications}~\cite{biem2010ibm}, and \emph{scientific workflows using provenance}~\cite{simmhan:sigmodrec:2005}.

\subsection{Stream Processing} Prior works~\cite{4620109,Repantis2006} explore the problem of composing streaming applications in a wide area P2P network, along with reuse of streams and tasks. Their DAG of tasks has an ontologically unique name for streams, 
newly submitted DAGs have their stream names matched against the existing streams, and identical streams are reused.  
Rather than just a lookup by stream names, we offer a more rigorous graph-based approach to distinctively identify equivalent tasks and their output streams. We also limit our work 
to a local cluster rather than wide area networks, 
and hence do not require the distributed probing mechanism they use to propagate state and connectivity. We can also use a centrally coordinate the reuse within the data center.  
Lastly, they do not adequately examine the removal of a submitted DAG -- as we saw, demerging can cause cascading impact on the deployed DAGs.

Reuse and sharing of queries has been explored for Distributed Stream Management Systems (DSMS), where tasks are 
query operators with  
well defined semantics that the system can take advantage.  
\cite{Zhou:2009:SDS:1687627.1687634} considers overlap between results of continuous queries and merges them into 
an equivalent query based on  
overlaps of attributes, predicates and streams. 
While we have similar approach for merging applications that share equivalent streams, we instead use a DAG model for comparing equivalence of typed tasks rather than require task semantics such as query behavior. This makes our work generalizable to any DAG-based streaming application.   

Others have also examined query admission, operator allocation and reuse as a set of inter-related problems that are solved as a constrained optimization problem~\cite{Kalyvianaki:2011:SSQ:2004686.2005583}. 
Their reuse of base (raw) stream and computed (derived) streams 
is similar to ours but they  
leverage knowledge of query operator semantics. They also impose 
resource constraints 
to restrict number of queries admitted into the system.  
We instead focus on opportunistic sharing of dataflow subsets to reduce their cost of execution rather than be constrained by a lack of resources. 
That said, DSPS execution on diverse edge-computing resources is an emerging research area for IoT and our techniques could be extended to the same~\cite{echo}.

\subsection{Scientific Workflows} Workflows have long been used to compose and publish e-Science applications through portals for loosely-coupled collaboration~\cite{4404805,atkinson2017scientific,myexperiment,opmw}. This approach can be replicated for sharing of streaming dataflows in the IoT domain as they grow popular. E.g., \emph{myExperiment}~\cite{myexperiment} provides a repository of workflows along with annotations and descriptions help locate, modify and reuse workflows. Reuse of the workflow composition is done manually, and modified workflows are usually published back for others to use. We instead consider reuse of running dataflows. 

Goderis, et al.~\cite{4032041} identifying similar worklows from existing DAGs based on their structure similarity. 
This is modeled as a  
subgraph isomorphism problem that is solved using existing techniques~
\cite{842269}. They also explore the problem of ranking the matched workflows.  
Similarly, our applications are also DAGs and we too offer techniques for  
graph structure matching but require exact matches of ancestor graphs to guarantee task equivalence.  
Others have performed statistical analysis on workflows from myExperiment 
to examine 
the reuse among workflows and the recurring set of services (tasks)~\cite{5569056}. 
Network analytics is then used for recommending services for new workflows being composed. We too leverage dataflow subset equivalence for reuse, but for running applications rather than for composing future ones.  
%
Provenance~\cite{simmhan:sigmodrec:2005} is metadata that captures the workflow execution trace to help users to decide if its generated outputs can be used, in part or in full, for their own workflow without performing a full execution~\cite{davidson2008provenance}. There are also mechanisms to efficiently search such traces to determine the appropriate dataset to reuse~\cite{anand2010techniques}. 

Unlike continuous stream processing, workflows execute in batch and generate files that are persisted. Hence, its the workflow composition that is reused for future executions rather the running workflow. The data products generated by prior workflow runs are also reused, with provenance as an enabler. 
Our focus instead is on reusing an actively running dataflow, with tasks added and removed during de/merge. There has also been limited research on using provenance for streaming dataflows~\cite{lim2009research,vijayakumar2006towards,wickramaarachchi2013continuous}, and it could offer an alternative approach to locate streams and equivalent tasks.

\section{Conclusion}
\label{sec:conclusion}
In this article, we have motivated the need and opportunity for reusing partial subset of tasks from streaming dataflows, in an emerging domain like IoT where data stream and dataflow sharing is expected to grow. We have formalized the problem definition rigorously with tight specifications on when tasks are equivalent between two dataflows, allowing them to be reused. We also offer invariants that will ensure that output consistency and resource minimization are achieved. We use these specifications to design merge and demerge algorithms for dataflows that are submitted and removed from the streaming system. We also map these algorithms to an implementation for the Apache Storm DSPS.

The algorithms are validated using a collection of real DAG structures from diverse science disciplines, hosted publicly at OPMW, and a smaller collection of real IoT streaming applications and their variants from RIoTBench. We empirically evaluate our merge and demerge algorithms by running the dataflows on a Storm commodity cluster, for sequential and random walk traces. 
For all the workloads, we see the expected drop in running task count using our Reuse strategy with a corresponding decrease in CPU resource usage, with up to $51\%$ reduction in cost. This makes our algorithms viable for deployment in collaboratory IoT environments.

As future work, we propose to examine the impact on DAG latency due to the indirection through the broker. We will also examine when to perform defragmentation and measure its impact on the application disruption, and the improvements in resource usage and latency. Lastly, it is also worth estimating the real-cost reduction on on-demand Cloud VMs, and techniques for fair billing of resources to the dataflows from different users that are being reused.





\bibliographystyle{IEEEtran}
\bibliography{paper}

\end{document}